\documentclass[
aps,
pra,
longbibliography,
superscriptaddress,
 amsmath,amssymb,
twocolumn,
]{revtex4-1}

\usepackage{graphicx}
\usepackage{dcolumn}
\usepackage{bm}
\usepackage{upgreek}
\usepackage{color}
\usepackage{hyperref}
\DeclareMathAlphabet{\mathpzc}{OT1}{pzc}{m}{it}
\usepackage{enumitem}   

\newcommand{\RR}{\right}
\newcommand{\LL}{\left}
\newcommand{\m}{\mathrm}
\newcommand{\dg}{\dagger}
\newcommand{\eref}[1]{Eq.~(\ref{#1})}
\newcommand{\fref}[1]{Fig.~\ref{#1}}
\newcommand{\puoli}{\frac{1}{2}}

\usepackage[normalem]{ulem}

\setlist[enumerate]{itemsep=-1mm}

\begin{document}

\title{Fast feedback control of mechanical motion using circuit optomechanics}

\author{Cheng Wang}
\affiliation{QTF Centre of Excellence, Department of Applied Physics, Aalto University, FI-00076 Aalto, Finland}

\author{Louise Banniard}
\affiliation{QTF Centre of Excellence, Department of Applied Physics, Aalto University, FI-00076 Aalto, Finland}

\author{Laure Mercier de L\'epinay}
\affiliation{QTF Centre of Excellence, Department of Applied Physics, Aalto University, FI-00076 Aalto, Finland}

\author{Mika A. Sillanp\"a\"a}
 \email{Mika.Sillanpaa@aalto.fi}
\affiliation{QTF Centre of Excellence, Department of Applied Physics, Aalto University, FI-00076 Aalto, Finland}

\date{\today}

\begin{abstract}
Measurement-based control, utilizing an active feedback loop, is a standard tool in technology. Feedback control is also emerging as a useful and fundamental tool in quantum technology and in related fundamental studies, where it can be used to prepare and stabilize pure quantum states in various quantum systems. Feedback-cooling of center-of-mass micromechanical oscillators, which typically exhibit a high thermal noise far above the quantum regime has been particularly actively studied and has recently been shown to allow for ground-state cooling using optical measurements. Here, we realize measurement-based feedback operations in an electromechanical system, cooling the mechanical thermal noise down to 3 quanta, limited by added amplifier noise. Counter-intuitively, we also obtain significant cooling when the system is pumped at the blue optomechanical sideband, where the system is unstable without feedback.
\end{abstract}

\maketitle


\section{Introduction}

In cavity optomechanics, quantum control of mechanical motion can be achieved via radiation pressure force from an optical light field on a mechanical degree of freedom in two different ways. Coherent quantum control involves applying a coherent pump tone to induce a strong coupling between the motion and an effective cold bath,  so that the combined system evolves to a desired state. In measurement-based feedback control, an error signal obtained from a measurement result is applied as a force on the mechanical oscillator through a time-delayed and carefully filtered feedback loop to steer and control the evolution of motional states.

Feedback control and its ability to achieve cooling of massive mechanical objects has been investigated earlier both theoretically and experimentally. It was first demonstrated in optics \cite{Cohadon1999}, with active experimental research following along the same lines \cite{Bouwmeester2006,Poggio2007FB,Vinante2008FB,Kippenberg2015FB,Bowen2016sqCool,Vitali2017FB,Kippenberg2017SquA,Becker2018symph,Groblacher2019FB,Muhonen2022FB}. Recently, feedback cooling down to the ground state was achieved for an ultrahigh quality factor SiN membrane resonator \cite{Schliesser2018FB}.
Feedback cooling also allowed to bring a 10 kg mass in the Advanced LIGO gravitational-wave detector close to its motional ground state \cite{LIGO2009, LIGO2021}. Besides massive oscillators, levitated nanoparticles have been successfully feedback-cooled \cite{Novotny2012FB,Paternostro2017FB,Raizen2011Levit,Quidant2019FB,Iwasaki2019FB}, some recent experiments even reaching the motional ground state \cite{Aspelmeyer2020Levit,Novotny2021FB}.


Feedback control applied to a microwave optomechanical system \cite{Lehnert2008Nph,Teufel2011b} has yet to be realized. The implementation poses experimental challenges, but also carries potential for operation deep in the quantum regime for which electromechanical systems are generally well suited. Typical micro-fabricated electromechanical resonators have rather high frequencies ($>5\,\rm MHz$), which sets constraints for a digital realization of a control system, as high processing rates are required. Furthermore, since the electromagnetic degree of freedom has to react fast to the control, a microwave cavity with a high external coupling is necessary. This poses further challenges, as large external couplings are not easily combined with mechanical elements directly integrated in superconducting on-chip cavities. Here, we realize feedback control in an electromechanical system employing a drum mechanical membrane, using a scheme adapted for this system, where we use a coherent tone to carry out a strong measurement, and a modulated tone to apply the adequate feedback force on the system.


\section{Theory}

In measurement-based feedback cooling of a mechanical oscillator, the motion is continuously monitored with a very high precision which allows to derive the oscillator's speed. A force proportional to the speed, which therefore acts as a viscous force, is fed back to the oscillator. This force artificially damps the motion without adding the fluctuation counterpart usually linked to damping mechanisms. This reduces the displacement variance, so that the oscillator is effectively cooled. In order to cool the oscillator's thermal occupancy near the quantum ground state, it is critical that the measurement is close to the quantum-limited sensitivity. Indeed, a measurement noise higher than the level of position quantum fluctuations results in a feedback force noise limiting the cooling efficiency above the quantum ground state.


\subsection{Basic principle of feedback cooling}

The principle of feedback cooling of mechanical oscillations is fairly well-known \cite{Tombesi1998FB,Vitali2002FB,Schwab2003FB,Aspelmeyer2008Cool,Nori2017FBreview,Genes2019FB} and is recalled here only briefly. The position $x$ (for the moment given in meters) of an oscillator of mass $M$, frequency $\omega_m$ and damping rate $\gamma$ follow the evolution equation:
\begin{equation}
\ddot{x}(t) = -\omega_m^2 x(t) - \gamma \dot{x}(t) + F_{\rm th}(t)/M,
\end{equation}
where $F_{\rm th}(t)$ is the Langevin force whose spectrum $S_F[\omega]$ in the classical limit is $S^{th}_F[\omega]=2k_BTM\gamma$, with $T$ the temperature of the oscillator's bath and $k_B$ the Boltzmann constant. The spectrum of the position of the free oscillator is
\begin{equation}
\label{eq:SxnoFB}
S_x[\omega]= \frac{2k_BT\gamma}{M[(\omega_m^2-\omega^2)^2+\gamma^2\omega^2]} \,.
\end{equation}
The damping rate appears both in the intensity of the coupling to the thermal bath and as the bandwidth of the Lorentzian mechanical spectrum. Applying a damping feedback force of $F_{\m{FB}} = -gM\gamma \dot{x}$, where $g$ is the feedback gain, broadens the spectrum $\gamma \rightarrow \gamma(1+g)$, resulting in:
\begin{equation}
\label{eq:SxFBgen}
S_x[\omega]= \frac{2k_BT\gamma}{M[(\omega_m^2-\omega^2)^2+\gamma^2(1+g)^2\omega^2]} \,.
\end{equation}
This process of damping the oscillator without adding fluctuations has been coined cold damping. Effectively, the resulting spectrum is that of an oscillator of damping $\gamma(1+g)$ at a temperature $T/(1+g)$, lower than the temperature of the bath $T$. This effect is therefore also called feedback cooling. The process requires a detection and a reaction on the measurement result much faster than the decay time of the oscillator $1/\gamma$. The cooling efficiency is limited by the amount of noise added in the feedback loop. This noise predominantly comes from the background noise in the detection of the oscillator's position.



\subsection{Microwave optomechanical detection}

We consider an archetypal optomechanical system, where a single mechanical harmonic mode (frequency $\omega_m$, and damping rate $\gamma$) is coupled to an electromagnetic cavity (frequency $\omega_c$, and damping rate $\kappa$). We ignore internal losses of the cavity. The cavity is probed by a strong coherent field (frequency $\omega_p$) which is detuned  from $\omega_c$ by the amount of $\Delta =\omega_p - \omega_c$. The probing induces an effective optomechanical coupling $G=g_0\sqrt{n_c}$, where $n_c$ is the number of photons driven in the cavity by the tone, and $g_0$ is the vacuum optomechanical coupling.

In order to describe the feedback process, we treat the system using the standard input-output theory of optical cavities. Now we treat the mechanical oscillator with dimensionless position $x(t)$ and momentum $p(t)$. We also define the dimensionless quadratures of the field in the cavity, $x_c(t)$ and $y_c(t)$ in the frame rotating at the cavity frequency. The equations of motion in the frequency domain are
\begin{equation}
\label{eqmot}
\begin{split}
\chi_c^{-1} x_c &= - \Delta y_c +  \sqrt{\kappa} x_\mathrm{c,in} \,, \\
\chi_c^{-1} y_c &= \Delta x_c - 2  G x + \sqrt{\kappa} y_\mathrm{c,in} \,,\\
- i \omega x &= \omega_m p \,, \\
\left(\gamma- i \omega\right)  p &= -  \omega_m x - 2  G  x_c + \frac{f_{\rm th}}{\omega_m} + \frac{f_{\rm fb}}{\omega_m}\,. \\
\end{split}
\end{equation}

Here, the cavity susceptibility is $\chi_c^{-1} = \frac{\kappa}{2} - i \omega$, and $x_\mathrm{c,in}$ and $y_\mathrm{c,in}$ are the input noise operators for the cavity. Finally, $f_{\rm th}$ and $f_{\rm fb}$ are scaled forces: $f_{X} = \omega_m F_{X}/(Mx_{\rm zpf})$, where $F_X$ is a force, $M$ the effective mass and $x_{\rm zpf}$ the zero-point fluctuations of the oscillator.

The feedback force $f_{\m{fb}}$ is now present in addition to the thermal force. If information on the measured observable, position $x$, is contained in the feedback force, a closed feedback loop is formed.

In a generic optomechanical measurement, the output field
\begin{equation}
\label{eq:youtgen}
\begin{split}
y_{\m{out}} =  \sqrt{\kappa} y_c -  y_{\m{c,in}}
 \end{split}
\end{equation}
emitted from the cavity carries information about $x$. The feedback force nature is designed by choosing a suitable processing, or filter function, to the measured $y_{\m{out}}$. In a real situation, the measurement can only provide an approximation of $x$. This is primarily because a significant amount of noise is added to $y_{\m{out}}$ before it is converted into a force. In microwave experiments, this noise, denoted as a random field $y_{\rm add}(t)$, is due to transistor amplifiers. Even in the best cases, this noise is at least an order of magnitude higher than the quantum limit. The amplifier noise is typically characterized as the added number of noise quanta $\langle y_{\rm add}[\omega] y_{\rm add}[-\omega] \rangle = n_{\m{add}}$. 

\begin{figure*}[t]
  \begin{center}
   \includegraphics[width=0.99\textwidth]{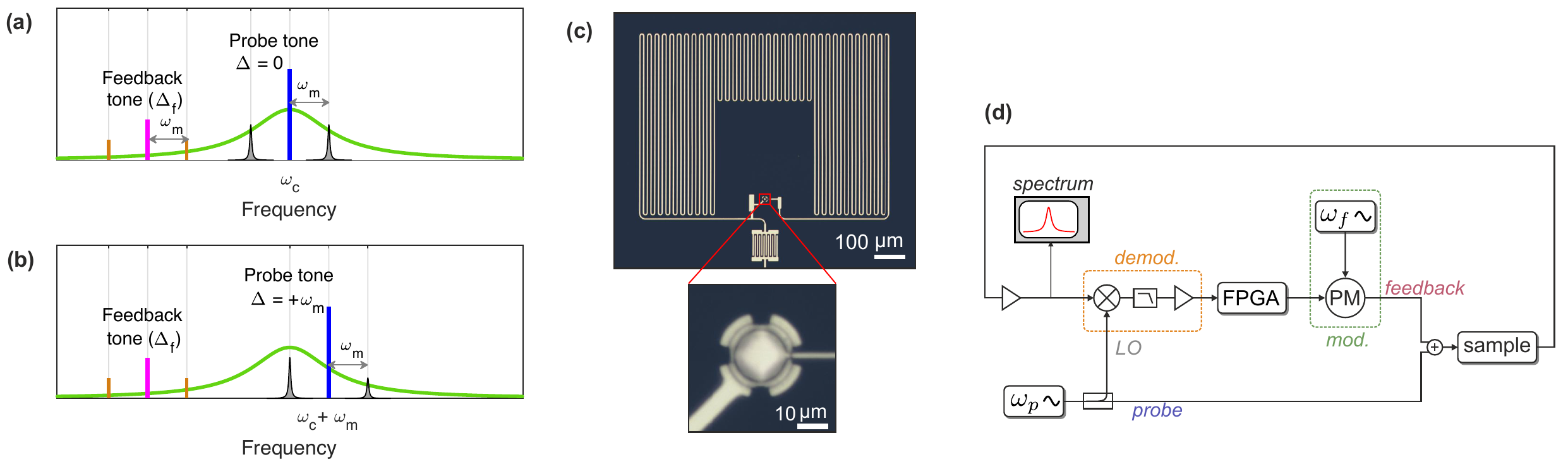}
    \caption{\textit{Feedback setup.} (a) The basic frequency scheme of feedback cooling in cavity optomechanics, with the strong probe tone set at the cavity frequency ($\Delta = 0$). (b) The probe tone set alternatively to the blue mechanical sideband ($\Delta = \omega_m$). (c) Optical micrograph a similar circuit-electromechanical device. The aluminum drumhead oscillator of diameter 13 $\mu$m is connected to a meander inductor to form a cavity strongly coupled to a transmission line through a large external finger capacitor. A zoom on the area of the drumhead is indicated with red dashed lines. (d) Simplified schematic of the microwave circuit around the electromechanical device; PM means phase modulator.
    } 
    \label{fig:figure1}
 \end{center}
\end{figure*}

The feedback force is obtained by applying a filter function  $A[\omega]$ to the signal, including a gain, and scaled for convenience by $\sqrt{\kappa}$:
\begin{equation}
\label{eq:FFBgen}
\begin{split}
f_{\m{fb}}[\omega] = \frac{A[\omega]}{\sqrt{\kappa}} \left(y_{\rm out}[\omega] + y_{\rm add}[\omega] \right) \,.
\end{split}
\end{equation}

To approximate a force proportional to the oscillator's velocity, we take the filter function $A[\omega]$ to be a phase-shifting (phase-shift $\phi\in \mathbb{R}$) and amplifying (gain $A_0>0$) application
\begin{equation}
\label{eq:transfunc}
\begin{split}
A[\omega] = A_0  \exp \left(- i \phi\frac{  \omega}{\omega_m} \right) \,.
 \end{split}
\end{equation}
%


\subsubsection{Resonant probing}

The unresolved sideband (bad-cavity) situation $\kappa \gg \omega_m$, where the cavity follows the mechanics without delay, allows for a simple treatment of the entire process. Our experimental parameters, where $\kappa \approx \omega_m$, do not well satisfy this condition. The basic case of resonant probing ($\Delta = 0$), as shown in \fref{fig:figure1} (a), allows for some analytical results at arbitrary sideband resolution. Here, the effective susceptibility of the oscillator is similar to that implied by \eref{eq:SxFBgen} 
\begin{equation}
\label{eq:chifb}
\begin{split}
\chi_{\m{fb}}[\omega] & = \frac{1}{-i \omega  \gamma_{\m{eff}}  +\omega_{\m{eff}}^2 - \omega^2 } \,.
\end{split}
\end{equation}
Here, for large mechanical quality factors $\omega_m/\gamma\gg 1$ (in the experiment $\omega_m/\gamma\sim 10^5$), the effective mechanical frequency and damping rate are, respectively
\begin{subequations}
\begin{alignat}{2}
& \omega_{\m{eff}} \simeq \omega_m + \frac{2G A_0 \left(\kappa \cos \phi + 2 \omega_m \sin \phi\right)}{\kappa^2+4 \omega_m^2}\,, \label{eq:omgeff}\\
& \gamma_{\m{eff}} \simeq \gamma + \frac{4 G A_0 (\kappa \sin \phi-2 \omega_m \cos \phi )}{\kappa^2+4 \omega_m^2} \,.
\label{eq:gammaeff}
\end{alignat}
\end{subequations}
%
At the optimum feedback phase satisfying
\begin{equation}
\phi_m =\tan^{-1} \LL( -\frac{\kappa}{2\omega_m} \RR)+\pi
\label{eq:phimax}
\end{equation}
the resonant frequency is unchanged, and the damping is maximized, with the feedback-induced damping
\begin{equation}
 \gamma_{\m{fb}} =\frac{4 G A_0}{\sqrt{\kappa^2 + 4\omega_m^2} } \,.
\label{eq:gammafb}
\end{equation}
The oscillator is supposed to couple to a bath with a thermal occupation number $n_m^T$, which is usually much larger than one. As mentioned earlier, the added feedback damping will induce cooling of the oscillator. However, there are competing processes which limit the cooling effect. 

The mechanical noise energy is obtained from the spectral density of the oscillator's displacement, where we can identify three contributions. The thermal plus zero-point fluctuation spectrum is cooled via the enhanced damping down to the variance
\begin{equation}
\begin{split}
 n_T & = \frac{\gamma}{\gamma_{\m{eff}}} \LL(n_m^T + \puoli \RR) \,.
\end{split}
\end{equation}
%
As in generic optomechanical position measurements, the quantum backaction of the measurement tends to heat up the oscillator linearly with the cooperativity, adding a mechanical population
\begin{equation}
\label{eq:nqba}
\begin{split}
n_{\m{qba}} = C_{\m{eff}} \frac{\kappa^2}{\kappa^2 + 4 \omega_m^2} \,.
\end{split}
\end{equation}
The cooperativity $C_{\rm eff}$ appearing in this quantum backaction noise is the cooperativity defined from the damped oscillator's parameters
\begin{equation}
\label{eq:Ceff}
\begin{split}
C_{\m{eff}} = \frac{4G^2}{\kappa \gamma_{\m{eff}}} \,.
\end{split}
\end{equation}
The increased damping of the oscillator (reduced $C_{\m{eff}}$) thus makes the oscillator less susceptible to the quantum backaction, and the backaction contribution $n_{\m{qba}}$ is reduced with increasing feedback gain.

The background noise in the detection in typical microwave-optomechanical systems is dominated by the microwave amplifier noise $n_{\rm add}$. Here, it is fed back to the oscillator, leading to additional mechanical fluctuations. Another, more fundamental contribution is due to vacuum fluctuations associated to the measurement, which are also fed back to the mechanical oscillator. The sum of these is
\begin{equation}
\label{eq:nFB}
\begin{split}
n_{\rm{fb}} = \frac{A_0^2}{2 \kappa \gamma_{\m{eff}} } \LL(n_{\m{add}} + \puoli \RR)  \,.
\end{split}
\end{equation}
The total remaining mechanical occupation $n_m$ under feedback cooling satisfies 
\begin{equation}
\label{eq:nmFB}
\begin{split}
n_m +\puoli= n_T + n_{\m{qba}} + n_{\m{fb}}  \,,
\end{split}
\end{equation}
which decreases with increasing gain $A_0$, then, for high gain, starts increasing again as $n_{\rm fb}$ becomes the dominant contribution to the occupation. The optimum cooling is reached when the oscillator is strongly damped, $n_T \ll 1$, and when the contributions of backaction and noise injection balance each other. This occurs when $\frac{G}{A_0} = \frac{1}{4} \sqrt{1+2 n_{\m{add}}} \sqrt{\kappa^2+4\omega_m^2}/\kappa$, and results in the optimum cooling
\begin{equation}
\label{eq:nmmin}
\begin{split}
n_{m,\m{min}} + \puoli= \puoli \sqrt{1+2 n_{\m{add}}}  \,.
\end{split}
\end{equation}
In order to cool down to the ground state with ${n_m < 1}$, one has to reach $n_{\m{add}} < 4$. This is beyond reach of transistor amplifiers, but is possible with near-quantum limited Josephson parametric amplifiers.

We now discuss the detection aspect. Under resonant cavity probing, the quadrature of the cavity output $y_{\rm out}=i(a^\dagger_{\rm out}+a_{\rm out})/\sqrt{2}$, where $a_{\rm out}$ is the output field annihilation operator, displays a mechanical signature at the optomechanical sidebands around the probe tone as shown by \eref{eq:youtgen}. In the experiment, the signal used to establish the feedback loop is the demodulated signal coming out of the cavity. For demonstration purposes, we also record the heterodyne spectrum
\begin{equation}
\label{eq:Sout}
\begin{split}
S_{\m{out}}[\omega] = \langle a^
 \dg_{\m{out}}[\omega] a_{\m{out}}[-\omega] \rangle + \puoli \,.
\end{split}
\end{equation}
%
Inference of the state of the mechanical oscillator based on the in-loop spectrum is complicated by the fact that the injected and reflected noises are out-of-phase, which leads to ``noise squashing", or destructive interference close to the mechanical resonance. This becomes relevant at strong feedback strength around the optimum cooling, \eref{eq:nmmin}. In the bad-cavity case, one can easily identify correlations leading to the squashing, see Appendix \ref{sec:badcavityresults}. In the case of arbitrary sideband resolution, we calculate the theoretical output spectra numerically (see Appendix \ref{sec:fbnumerics}) from the solution of the equations of motion, \eref{eqmot}, and from the input noise correlators.

\subsubsection{Blue-sideband probing}

For optomechanical systems with $\kappa \lesssim \omega_m$, blue-sideband probing leads to optomechanical antidamping. The damping rate is reduced by
\begin{equation}
\label{eq:gammaopt}
\gamma_{\m{opt}} =  \frac{4 G^2}{\kappa} \frac{1}{1+ \LL(\frac{\kappa}{4 \omega_m} \RR)^2}\,.
\end{equation}
If this antidamping rate overcomes the intrinsic, or otherwise enhanced damping rate of the mechanical oscillator, the latter becomes unstable.

It has been shown that blue-sideband pumping can be utilized to drive a stable steady state, but only if combined with other processes that stabilize the system. Examples include optomechanical dissipative squeezing and entanglement \cite{WoolleyBAE,SchwabSqueeze,TeufelSqueeze,Squeeze,Entanglement}, which count on steadying dynamical backaction effects dominating backaction noise in some regimes, to allow for the squeezing of a quadrature, in spite of a simultaneous increase of the effective bath temperature for that quadrature because of backaction noise. 

A similar competition of effects exists in the present case: large probe powers provide large detection sensitivities beneficial to the feedback process, but also large optomechanical amplifications that can result in instability. The gain of the feedback loop can be chosen independently from the probe power, so that a good configuration of feedback parameters is expected to produce cooling. The question is then whether the strong cold damping effect obtained from this efficient measurement can, in some range of parameters, dominate optomechanical anti-damping. We find that the answer is yes. The effective mechanical damping rate in the scheme of \fref{fig:figure1} (b) is the sum of the rates due to dynamical backaction, and of the feedback cooling:
\begin{equation}
\label{eq:gammaeffBSB}
\gamma_{\m{eff}} = \gamma - \gamma_{\m{opt}} + \frac{4 G A_0 \left[ (\kappa^2 + 8 \omega_m^2) \sin \phi  -2 \kappa \omega_m \cos \phi\right] }{\kappa^3 + 16 \kappa \omega_m^2} \,.
\end{equation}

Because the feedback has to first counteract the amplification induced by dynamical backaction, a larger gain as compared to resonant probing is needed, which in turn will inject more noise and tend to reduce the cooling performance. Ground-state cooling is still possible, but very little added noise can be tolerated.


\begin{figure*}[t]
  \begin{center}
   {\includegraphics[width=0.85\textwidth]{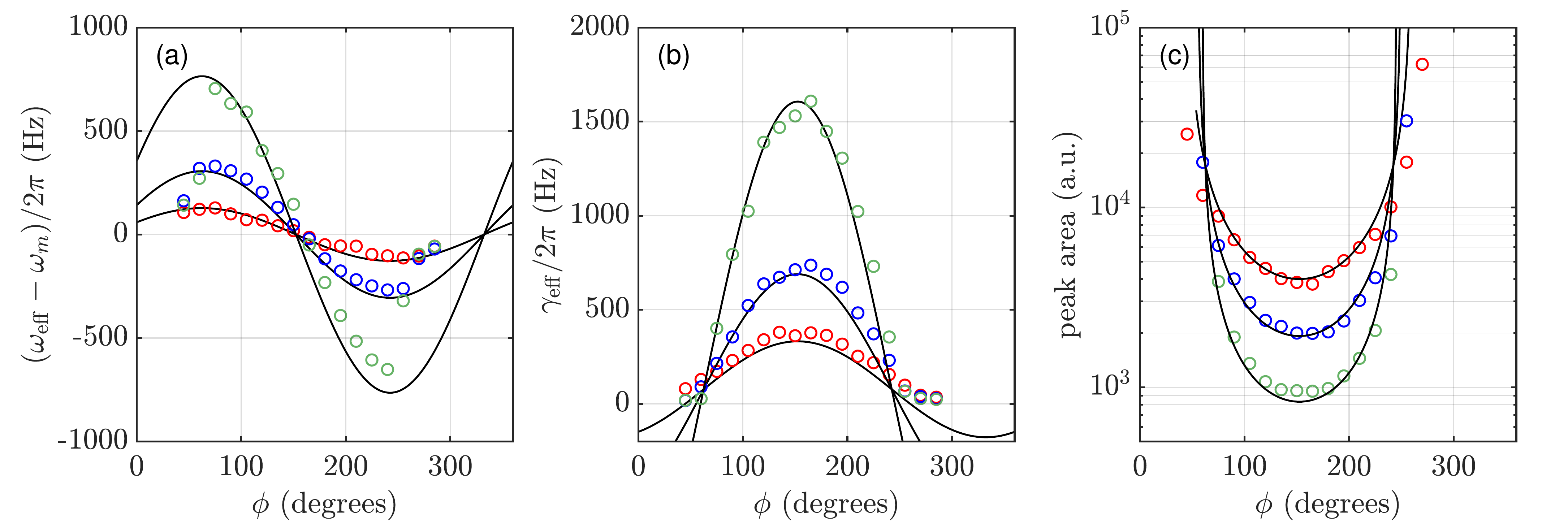} } 
    \caption{\textit{Feedback control measurements}. We use a strong probe tone at the cavity center frequency ($\Delta=0$), with the effective coupling $G/2\pi \simeq 420$ kHz. The data in each panel correspond to the feedback gain $A_0/2\pi = 2.8$ kHz (red), $A_0/2\pi = 6.7$ kHz (blue), and $A_0/2\pi = 16.7$ kHz (green). The solid lines are theoretical fits. (a) Mechanical frequency shift, and (b) effective damping rate, as functions of feedback loop phase. (c) The area of the mechanical peak in the heterodyne output spectrum.}
    \label{fig:figphase}
 \end{center}
\end{figure*}

\subsection{Feedback scheme}

In some feedback-cooling experiments, a single laser is used both for probing and applying the feedback force by radiation pressure \cite{Vitali2017FB,Muhonen2022FB}. 
A situation described by \eref{eqmot} requires a separate method to create the force. One possibility is to use direct mechanical actuation \cite{LIGO2009}. Most optomechanical experiments have utilized another laser dedicated to applying the feedback force. We adapt this technique in this work, and create the feedback force by suitable modulation of another microwave tone, a relatively weak feedback tone.

The basic frequency scheme is shown in \fref{fig:figure1} (a), where the probe tone frequency is set at the resonance of the cavity. The feedback tone is typically detuned from the cavity by a detuning $\Delta_f$ larger than several cavity linewidths. This large frequency separation between microwave tones allows to treat all optomechanical processes independently. Similar reasoning holds also for the other explored alternative, the blue-sideband probing shown in \fref{fig:figure1} (b).

%

%
At room temperature, the cavity output field is homodyne-detected
by demodulating with a local oscillator at the probe tone frequency $\omega_p$, as seen in \fref{fig:figure1} (d). The phase of the local oscillator (with respect to the phase of the driving tone) is tuned to measure the quadrature $y_c$ carrying the most information on the mechanical oscillator. 
The demodulated quadrature is a direct record of the position, appearing as an oscillatory signal at the frequency $\omega_m$ in the laboratory frame. This signal, once phase-shifted and amplified, is used to realize a weak phase modulation of a second microwave tone (the feedback tone) at the frequency $\omega_f=\omega_c+\Delta_f$. This effectively generates a triplet of frequencies separated by $\omega_m$, centered at $\omega_f$. With an adequate setting for the feedback phase-shift, the amplitude of each sideband of the driving triplet is approximately proportional to the velocity $\dot{x}$ while the central peak has a constant, much larger amplitude. Due to the nonlinearity of the optomechanical interaction, each sideband interferes with the central peak to produce a feedback force linear in sideband amplitude, that is, proportional to $\dot{x}$. Cross-products of sidebands generate a force dependent on the mechanical energy, which is maintained negligible in comparison to the linear feedback force by keeping the amplitude of the central peak much larger than the amplitude of the sidebands.

\begin{figure*}[t]
  \begin{center}
  {\includegraphics[width= 0.99\textwidth]{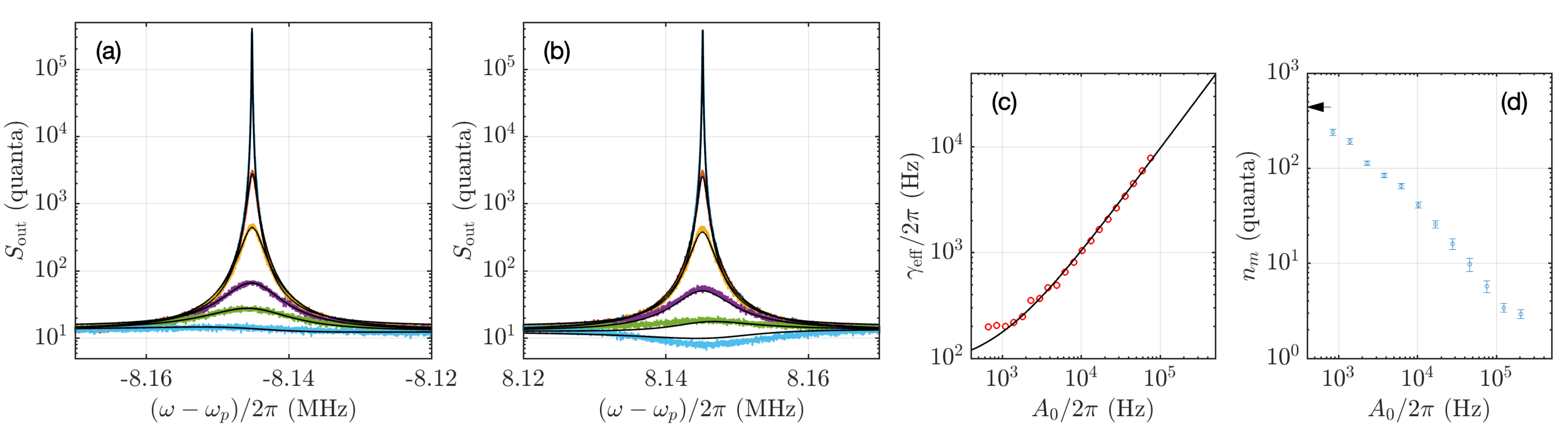} }
    \caption{\textit{Feedback cooling.} The parameter values are $G/2\pi \simeq 427$ kHz, $\phi \simeq 143$ degrees. (a), (b) In-loop heterodyne output spectra at the lower and upper sideband, respectively. The gain values were $A_0/2\pi \simeq [0, 10, 28, 76, 125, 206]$ kHz from top to bottom. The solid lines are theoretical fits. (c) Damping rate extracted by fitting Lorentzian curves to the spectra at lower gain values, together with a fit to \eref{eq:gammaeff}. The data is shown in the range where the peaks are roughly Lorentzian. (d) Mechanical occupation as a function of feedback gain. The horizontal arrow indicates the value $n_m \simeq 440$ at zero gain.}
    \label{fig:cooling}
 \end{center}
\end{figure*}

\section{Experimental setup}

\subsection{Electromechanical device} 

A microwave optomechanical device is used in which an aluminium drum oscillator is coupled to a superconducting microwave resonator (the cavity). Photographs of the device are shown in \fref{fig:figure1} (c). The aluminum drum oscillator is a parallel plate capacitor with a vacuum gap, consisting of an aluminium membrane suspended above an aluminum electrode. It oscillates at a frequency $\omega_m/2\pi = 8.14$ MHz and has an intrinsic damping rate $\gamma/2\pi= 76$ Hz. 
An LC circuit formed by this plate capacitor and a meandering inductor sustains a resonance at a microwave frequency $\omega_c/2\pi = 5.35$ GHz. The microwave cavity is strongly coupled to a transmission line thanks to a large interdigitated capacitor. The cavity is overcoupled, with a decay rate $\kappa/2\pi= 8.5$ MHz largely dominated by external coupling.

The device is maintained at a somewhat elevated temperature $80$ mK, where $n_m^T \simeq 205$, in a dilution refrigerator. We chose an operating temperature clearly higher than the base temperature, because we observed intermittent ``spiking" \cite{Collin2019demag} of the mode temperature at lower temperatures. 

The single-photon optomechanical coupling is found to be $g_0/2\pi\simeq 130\,\rm Hz$. The calibration of the effective coupling $G$ of the probing tone is realized by monitoring the sideband-cooling effect as described in Appendix \ref{app:calibrations}. The values of the enhanced couplings of all tones used in the work are inferred from this calibration and from the measurement of the cavity susceptibility.

\subsection{Feedback setup} 

The cavity output signal, \eref{eq:youtgen}, which mainly consists of the two optomechanical sidebands generated by the probe tone, is amplified inside the refrigerator with an amplifier exhibiting the effective noise $n_{\m{add}} \simeq 13$ quanta, then demodulated using the probe tone as a local oscillator to realize a homodyne detection. The demodulation result is passed through an analog band-pass filter to retain the signal oscillating at the mechanical frequency, while limiting unnecessary broadband noise before digitization. The signal is then sent to a 14-bit FPGA-backed acquisition device (Red Pitaya STEMlab 125-14) capable of 125 MHz input and output sampling rate. The FPGA card is programmed to replay this signal, after further digital band-pass filtering, time delaying, and amplification. 

The FPGA output signal is sent to a microwave phase modulator to modulate the feedback tone at the frequency $\omega_f$. This phase modulation, being very weak, is essentially comparable to an amplitude modulation, and generates mechanical-momentum-dependent sidebands $\pm \omega_m$ around a strong coherent peak at $\omega_f$. The details are given in Appendix \ref{app:emforce}.

Since the feedback tone (and its modulation sidebands) sits far on the red side of the cavity, the cavity susceptibility significantly shifts the phase of the feedback force, by an estimated $72^\circ$. The tunable contribution to the phase shift is eventually tuned to produce a total phase shift of $\phi_m + 2n\pi$ ($n\in {\mathbb{N}}$) between the position signal and the feedback force. As long as the total number of additional periods $n$ by which the force is delayed from the momentum signal remains much smaller than the quality factor of the mechanical oscillator, the feedback quality is not significantly affected by this additional delay.

Each sideband of the feedback triplet, by interfering within the nonlinear optomechanical interaction with the strong central coherent peak of the triplet, generates a term in the feedback force with a slightly different phase-shift, as the cavity susceptibility and microwave transmission lines contribute differently to the total phase shift for each component of the driving triplet. These two contributions to the feedback force containing versions of the mechanical signal phase-shifted by different amounts do not necessarily add up fully constructively (see Appendix \ref{app:emforce}), but one is sufficiently attenuated by the cavity susceptibility to limit the effect of a destructive interference and allow for a relatively strong feedback force in the experimental situation.  

\section{Results}

\subsection{Probing at the cavity frequency}

The probe tone is first positioned at the cavity frequency ($\Delta = 0$), such that the displacement spectra of the oscillator is encoded in the cavity output spectrum with high efficiency while maintaining mechanical stability. The feedback tone is detuned by $\Delta_f/2\pi = -20$ MHz from the cavity frequency. This detuning is chosen such that $|\Delta_{f}| >2\omega_m$ to avoid integrating stray feedback field components into the measurement of the position, but is kept of the order of $\kappa$ to allow for a significant response of the cavity.



Next, we vary the feedback phase $\phi$ and record the properties of the mechanical peaks in the heterodyne spectrum. The peaks are primarily characterized by their frequency, \eref{eq:omgeff}, and linewidth, \eref{eq:gammaeff}. We show the phase dependence of these quantities in \fref{fig:figphase} (a), (b). The data is not plotted in the regime where the system is unstable ($\gamma_{\m{eff}} < 0$). Aside from the effect of  the feedback, the mechanical frequency is expected to undergo a red-shift under the strong probe driving due to the second-order optomechanical coupling. The shift is given by $\delta f_2 = -\frac{1}{2} g_2 n_c$, where $g_2 = \frac{1}{2}\frac{d^2 \omega_c}{d x^2} x_{\m{zpf}}^2$. With the experimental parameters, we expect $\delta f_2/2\pi \simeq 400$ Hz. However, the frequency at zero feedback gain is observed to be red-shifted by 1.2 kHz with respect to its value calibrated independently. We believe that the additional shift is due to occasional drifts of the intrinsic mechanical frequency observed during the cooldown. For the feedback-induced frequency shifts shown in \fref{fig:figphase} (a), we adjust the zero to correct for the uncertainty in the intrinsic frequency.

With this adjustment, we reach a good agreement with the theoretical predictions, letting the unknown phase shift due to the time delay in cables as another adjustable parameter. We also study phase dependence of the area of the mechanical peaks as shown in \fref{fig:figphase} (c). In the limit of weak feedback, where the noise squashing plays little role, the peak area is a good measure of the mechanical mode temperature. This condition is satisfied in the data shown in \fref{fig:figphase}, where $\gamma_{\m{eff}}$ reaches values up to $\sim 2 \pi \cdot 1$ kHz. A cooling by one order of magnitude is observed in \fref{fig:figphase} (c), where the reference is the crossing point of the curves corresponding to zero cooling or amplification.


The maximum damping from the phase sweeping measurements is determined at a phase value around $145^\circ \pm 5^\circ$, which corresponds to an optimized loop delay to a exert damping force proportional to velocity. Notice that this phase differs from the optimum value $\phi_m = 90^\circ$ expected in the extreme bad-cavity limit. With this optimized phase, we proceed to vary the feedback gain up to large values and investigate the maximum cooling we can achieve.

\begin{figure*}[t]
  \begin{center}
   {\includegraphics[width=0.99\textwidth]{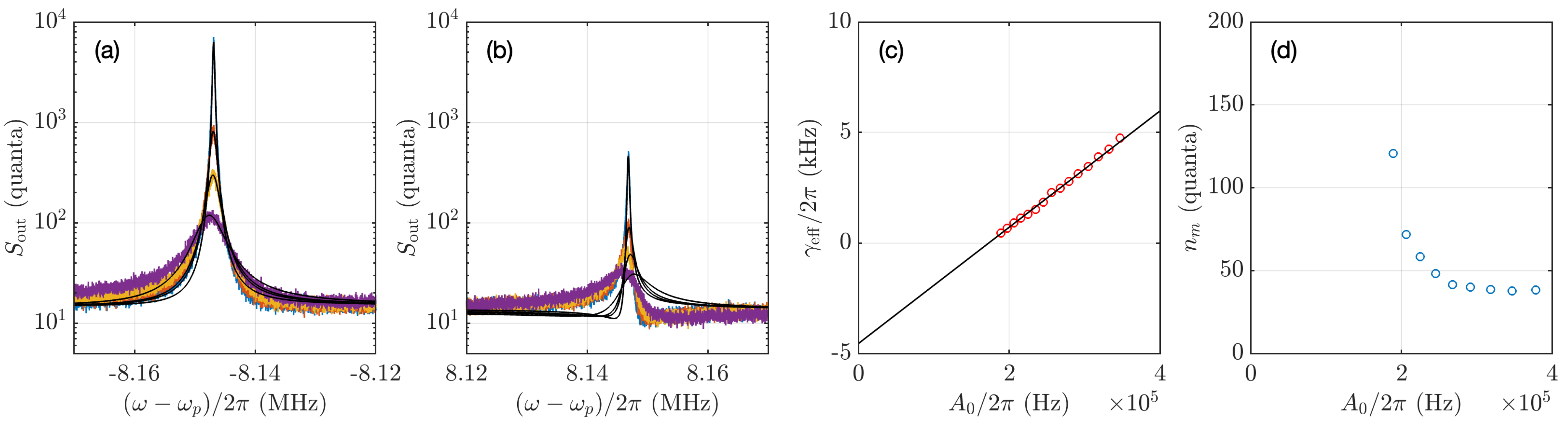} }
    \caption{\textit{Feedback stabilization and cooling of an intrinsically unstable system.} The probe tone is set at the blue sideband frequency ($\Delta = \omega_m$), and the effective coupling is $G/2\pi \simeq 104$ kHz. The gain values are $A_0/2\pi \simeq [189, 225, 267, 378]$ kHz from top to bottom. The panels present the same quantities as in the resonant pumping case, \fref{fig:cooling}; (a), (b) Lower and upper sideband peaks around the probe tone, respectively. (c) Effective damping in the stable range. (d) Mechanical occupation as a function of feedback gain.}
    \label{fig:bsb}
 \end{center}
\end{figure*}

We show in \fref{fig:cooling} (a), (b), the two peaks of the in-loop heterodyne spectra. At high gain, the peaks are not equal. The lower sideband of the probe tone exhibits less squashing than the upper sideband. This can be interpreted as a manifestation of the sideband asymmetry that has been studied in various optomechanical systems \cite{Schwab2014Asymm,Harris2015asym,Simmonds2015qb}, see \eref{eq:squashbadcav}. For the theoretical fits, we used the bath temperature for each gain as the only adjustable parameter because we anticipate the mechanical bath may be heated up as the gain increases. We indeed observe a heating of the bath from $n_m^T \simeq 200$ to $n_m^T \simeq 370$ between zero and the high feedback gain values. Overall, the theory portrays a good fit to the data. There is some discrepancy at the highest gain values in the upper sideband, which we believe may be due to some noise from the feedback tone starting to circulate in the system.

According to \eref{eq:gammaeff}, the mechanical linewidth evolves linearly with respect to the feedback gain. We test this basic property as shown in \fref{fig:cooling} (c). At high gain values, the peaks get distorted due to noise squashing, and this data is not considered in this situation. The solid line in \fref{fig:cooling} (c) is a linear fit that is the basis for creating the theoretical spectra in \fref{fig:cooling} (a), (b), according to Eqs.~(\ref{eq:Lcal},\ref{eq:Lcaluse}). Finally, in \fref{fig:cooling} (d) we show the mechanical occupancy that can be reached in our setup. The occupancy is calculated with \eref{eq:nmFB} using the calibrated quantities, and the bath temperature obtained from fitting the spectra. We reach an occupation of $n_m \simeq 2.9 \pm 0.3$ quanta, which is significantly close to the ground state, and is limited by the amplifier added noise injected back to the sample.

\subsection{Probing at the blue optomechanical sideband}

We then moved the probe tone from the cavity center to the blue optomechanical sideband frequency. Since the probe tone at its final intended power pushes the oscillator far into the instability regime, the feedback parameters were first optimized at a low probe power, only incrementally reaching higher powers by exploring small parameter ranges around the stability regime. 
Figure \ref{fig:bsb} shows the result of gain sweeps performed at the highest probe power used in this configuration, and at the optimal feedback phase. The investigation of small gains is precluded by instability at low feedback efficiencies.

The lower sideband peak shown in \fref{fig:bsb} (a), which in this case is located right at the cavity frequency, exhibits a good agreement with the numerically computed lineshape for each gain displayed as black solid lines in the panel. The upper sideband peak shown in \fref{fig:bsb} (b) is strongly suppressed by the unfavorable cavity susceptibility as this signal is detuned from the cavity by approximately twice the cavity linewidth. With the upper sideband, we acknowledge a discrepancy between the theoretical lineshapes also displayed in the panel. We again anticipate some additional noise is circulating in the feedback loop at frequencies near the upper sideband of the probe tone, inducing additional squashing effects. The calibrated height of the peaks, however, matches well with the predictions. 

Similar to resonant pumping case, we again use the mechanical bath temperature as a free parameter in the fits to the lineshapes. Here we find a very strong technical heating when the feedback gain increases: the mechanical bath heats up to $n_m^T \simeq 5\cdot 10^3$ phonons at the maximum gain $A_0/2\pi\simeq 400$ kHz. 

As shown in \fref{fig:bsb} (c), a sizable damping rate $\gamma_{\rm eff}/2\pi\simeq 5$ kHz can be obtained in the blue-sideband configuration. This damping is not only significant considering that the oscillator is anti-damped at zero gain, but it is also nearly two orders of magnitude larger than the intrinsic damping. Finally, in \fref{fig:bsb} (d) we display the mechanical occupation inferred in the situation using our numerical model. We obtain a modest cooling down to $n_m \simeq 38$, limited primarily by the technical heating.


\section{Conclusions}

In summary, we demonstrated feedback cooling in a microwave optomechanical system. We reached a mechanical occupation $n_m \simeq 3$ quanta in a 8 MHz membrane resonator. The cooling is limited by the added noise of the microwave amplifier. By introducing a much less noisier Josephson parametric amplifier \cite{Siddiqi2015Amp}, ground-state cooling in the present system, with the parameters used in this work, should be well within reach. This will open up possibilities for feedback-based preparation of more sophisticated states, such as squeezed states \cite{Marquardt2008Sq} through backaction-evading measurements.

\begin{acknowledgments} We would like to thank Juha Muhonen and Marton Gunyho for useful discussions. We acknowledge the facilities and technical support of Otaniemi research infrastructure for Micro and Nanotechnologies (OtaNano). This work was supported by the Academy of Finland (contracts 307757, 312057), and by the European Research Council (101019712). The work was performed as part of the Academy of Finland Centre of Excellence program (project 336810). We acknowledge funding from the European Union's Horizon 2020 research and innovation program under grant agreement 824109, the European Microkelvin Platform (EMP), and QuantERA II Programme (13352189). L. Mercier de Lépinay acknowledges funding from the Strategic Research Council at the Academy of Finland (Grant No. 338565).
\end{acknowledgments}


\begin{thebibliography}{40}%
\makeatletter
\providecommand \@ifxundefined [1]{%
 \@ifx{#1\undefined}
}%
\providecommand \@ifnum [1]{%
 \ifnum #1\expandafter \@firstoftwo
 \else \expandafter \@secondoftwo
 \fi
}%
\providecommand \@ifx [1]{%
 \ifx #1\expandafter \@firstoftwo
 \else \expandafter \@secondoftwo
 \fi
}%
\providecommand \natexlab [1]{#1}%
\providecommand \enquote  [1]{``#1''}%
\providecommand \bibnamefont  [1]{#1}%
\providecommand \bibfnamefont [1]{#1}%
\providecommand \citenamefont [1]{#1}%
\providecommand \href@noop [0]{\@secondoftwo}%
\providecommand \href [0]{\begingroup \@sanitize@url \@href}%
\providecommand \@href[1]{\@@startlink{#1}\@@href}%
\providecommand \@@href[1]{\endgroup#1\@@endlink}%
\providecommand \@sanitize@url [0]{\catcode `\\12\catcode `\$12\catcode
  `\&12\catcode `\#12\catcode `\^12\catcode `\_12\catcode `\%12\relax}%
\providecommand \@@startlink[1]{}%
\providecommand \@@endlink[0]{}%
\providecommand \url  [0]{\begingroup\@sanitize@url \@url }%
\providecommand \@url [1]{\endgroup\@href {#1}{\urlprefix }}%
\providecommand \urlprefix  [0]{URL }%
\providecommand \Eprint [0]{\href }%
\providecommand \doibase [0]{http://dx.doi.org/}%
\providecommand \selectlanguage [0]{\@gobble}%
\providecommand \bibinfo  [0]{\@secondoftwo}%
\providecommand \bibfield  [0]{\@secondoftwo}%
\providecommand \translation [1]{[#1]}%
\providecommand \BibitemOpen [0]{}%
\providecommand \bibitemStop [0]{}%
\providecommand \bibitemNoStop [0]{.\EOS\space}%
\providecommand \EOS [0]{\spacefactor3000\relax}%
\providecommand \BibitemShut  [1]{\csname bibitem#1\endcsname}%
\let\auto@bib@innerbib\@empty
\bibitem [{\citenamefont {Cohadon}\ \emph {et~al.}(1999)\citenamefont
  {Cohadon}, \citenamefont {Heidmann},\ and\ \citenamefont
  {Pinard}}]{Cohadon1999}%
  \BibitemOpen
  \bibfield  {author} {\bibinfo {author} {\bibfnamefont {P.~F.}\ \bibnamefont
  {Cohadon}}, \bibinfo {author} {\bibfnamefont {A.}~\bibnamefont {Heidmann}}, \
  and\ \bibinfo {author} {\bibfnamefont {M.}~\bibnamefont {Pinard}},\
  }\bibfield  {title} {\enquote {\bibinfo {title} {Cooling of a mirror by
  radiation pressure},}\ }\href@noop {} {\bibfield  {journal} {\bibinfo
  {journal} {Phys. Rev. Lett.}\ }\textbf {\bibinfo {volume} {83}},\ \bibinfo
  {pages} {3174--3177} (\bibinfo {year} {1999})}\BibitemShut {NoStop}%
\bibitem [{\citenamefont {Kleckner}\ and\ \citenamefont
  {Bouwmeester}(2006)}]{Bouwmeester2006}%
  \BibitemOpen
  \bibfield  {author} {\bibinfo {author} {\bibfnamefont {Dustin}\ \bibnamefont
  {Kleckner}}\ and\ \bibinfo {author} {\bibfnamefont {Dirk}\ \bibnamefont
  {Bouwmeester}},\ }\bibfield  {title} {\enquote {\bibinfo {title} {Sub-kelvin
  optical cooling of a micromechanical resonator},}\ }\href@noop {} {\bibfield
  {journal} {\bibinfo  {journal} {Nature}\ }\textbf {\bibinfo {volume} {444}},\
  \bibinfo {pages} {75--78} (\bibinfo {year} {2006})}\BibitemShut {NoStop}%
\bibitem [{\citenamefont {Poggio}\ \emph {et~al.}(2007)\citenamefont {Poggio},
  \citenamefont {Degen}, \citenamefont {Mamin},\ and\ \citenamefont
  {Rugar}}]{Poggio2007FB}%
  \BibitemOpen
  \bibfield  {author} {\bibinfo {author} {\bibfnamefont {M.}~\bibnamefont
  {Poggio}}, \bibinfo {author} {\bibfnamefont {C.~L.}\ \bibnamefont {Degen}},
  \bibinfo {author} {\bibfnamefont {H.~J.}\ \bibnamefont {Mamin}}, \ and\
  \bibinfo {author} {\bibfnamefont {D.}~\bibnamefont {Rugar}},\ }\bibfield
  {title} {\enquote {\bibinfo {title} {Feedback cooling of a cantilever's
  fundamental mode below 5 m{K}},}\ }\href@noop {} {\bibfield  {journal}
  {\bibinfo  {journal} {Phys. Rev. Lett.}\ }\textbf {\bibinfo {volume} {99}},\
  \bibinfo {pages} {017201} (\bibinfo {year} {2007})}\BibitemShut {NoStop}%
\bibitem [{\citenamefont {Vinante}\ \emph {et~al.}(2008)\citenamefont
  {Vinante}, \citenamefont {Bignotto}, \citenamefont {Bonaldi}, \citenamefont
  {Cerdonio}, \citenamefont {Conti}, \citenamefont {Falferi}, \citenamefont
  {Liguori}, \citenamefont {Longo}, \citenamefont {Mezzena}, \citenamefont
  {Ortolan}, \citenamefont {Prodi}, \citenamefont {Salemi}, \citenamefont
  {Taffarello}, \citenamefont {Vedovato}, \citenamefont {Vitale},\ and\
  \citenamefont {Zendri}}]{Vinante2008FB}%
  \BibitemOpen
  \bibfield  {author} {\bibinfo {author} {\bibfnamefont {A.}~\bibnamefont
  {Vinante}}, \bibinfo {author} {\bibfnamefont {M.}~\bibnamefont {Bignotto}},
  \bibinfo {author} {\bibfnamefont {M.}~\bibnamefont {Bonaldi}}, \bibinfo
  {author} {\bibfnamefont {M.}~\bibnamefont {Cerdonio}}, \bibinfo {author}
  {\bibfnamefont {L.}~\bibnamefont {Conti}}, \bibinfo {author} {\bibfnamefont
  {P.}~\bibnamefont {Falferi}}, \bibinfo {author} {\bibfnamefont
  {N.}~\bibnamefont {Liguori}}, \bibinfo {author} {\bibfnamefont
  {S.}~\bibnamefont {Longo}}, \bibinfo {author} {\bibfnamefont
  {R.}~\bibnamefont {Mezzena}}, \bibinfo {author} {\bibfnamefont
  {A.}~\bibnamefont {Ortolan}}, \bibinfo {author} {\bibfnamefont {G.~A.}\
  \bibnamefont {Prodi}}, \bibinfo {author} {\bibfnamefont {F.}~\bibnamefont
  {Salemi}}, \bibinfo {author} {\bibfnamefont {L.}~\bibnamefont {Taffarello}},
  \bibinfo {author} {\bibfnamefont {G.}~\bibnamefont {Vedovato}}, \bibinfo
  {author} {\bibfnamefont {S.}~\bibnamefont {Vitale}}, \ and\ \bibinfo {author}
  {\bibfnamefont {J.-P.}\ \bibnamefont {Zendri}},\ }\bibfield  {title}
  {\enquote {\bibinfo {title} {Feedback cooling of the normal modes of a
  massive electromechanical system to submillikelvin temperature},}\
  }\href@noop {} {\bibfield  {journal} {\bibinfo  {journal} {Phys. Rev. Lett.}\
  }\textbf {\bibinfo {volume} {101}},\ \bibinfo {pages} {033601} (\bibinfo
  {year} {2008})}\BibitemShut {NoStop}%
\bibitem [{\citenamefont {Wilson}\ \emph {et~al.}(2015)\citenamefont {Wilson},
  \citenamefont {Sudhir}, \citenamefont {Piro}, \citenamefont {Schilling},
  \citenamefont {Ghadimi},\ and\ \citenamefont
  {Kippenberg}}]{Kippenberg2015FB}%
  \BibitemOpen
  \bibfield  {author} {\bibinfo {author} {\bibfnamefont {D.~J.}\ \bibnamefont
  {Wilson}}, \bibinfo {author} {\bibfnamefont {V.}~\bibnamefont {Sudhir}},
  \bibinfo {author} {\bibfnamefont {N.}~\bibnamefont {Piro}}, \bibinfo {author}
  {\bibfnamefont {R.}~\bibnamefont {Schilling}}, \bibinfo {author}
  {\bibfnamefont {A.}~\bibnamefont {Ghadimi}}, \ and\ \bibinfo {author}
  {\bibfnamefont {T.~J.}\ \bibnamefont {Kippenberg}},\ }\bibfield  {title}
  {\enquote {\bibinfo {title} {Measurement-based control of a mechanical
  oscillator at its thermal decoherence rate},}\ }\href@noop {} {\bibfield
  {journal} {\bibinfo  {journal} {Nature}\ }\textbf {\bibinfo {volume} {524}},\
  \bibinfo {pages} {325--329} (\bibinfo {year} {2015})}\BibitemShut {NoStop}%
\bibitem [{\citenamefont {Sch{\"a}fermeier}\ \emph {et~al.}(2016)\citenamefont
  {Sch{\"a}fermeier}, \citenamefont {Kerdoncuff}, \citenamefont {Hoff},
  \citenamefont {Fu}, \citenamefont {Huck}, \citenamefont {Bilek},
  \citenamefont {Harris}, \citenamefont {Bowen}, \citenamefont {Gehring},\ and\
  \citenamefont {Andersen}}]{Bowen2016sqCool}%
  \BibitemOpen
  \bibfield  {author} {\bibinfo {author} {\bibfnamefont {Clemens}\ \bibnamefont
  {Sch{\"a}fermeier}}, \bibinfo {author} {\bibfnamefont {Hugo}\ \bibnamefont
  {Kerdoncuff}}, \bibinfo {author} {\bibfnamefont {Ulrich~B.}\ \bibnamefont
  {Hoff}}, \bibinfo {author} {\bibfnamefont {Hao}\ \bibnamefont {Fu}}, \bibinfo
  {author} {\bibfnamefont {Alexander}\ \bibnamefont {Huck}}, \bibinfo {author}
  {\bibfnamefont {Jan}\ \bibnamefont {Bilek}}, \bibinfo {author} {\bibfnamefont
  {Glen~I.}\ \bibnamefont {Harris}}, \bibinfo {author} {\bibfnamefont
  {Warwick~P.}\ \bibnamefont {Bowen}}, \bibinfo {author} {\bibfnamefont
  {Tobias}\ \bibnamefont {Gehring}}, \ and\ \bibinfo {author} {\bibfnamefont
  {Ulrik~L.}\ \bibnamefont {Andersen}},\ }\bibfield  {title} {\enquote
  {\bibinfo {title} {Quantum enhanced feedback cooling of a mechanical
  oscillator using nonclassical light},}\ }\href@noop {} {\bibfield  {journal}
  {\bibinfo  {journal} {Nature Communications}\ }\textbf {\bibinfo {volume}
  {7}},\ \bibinfo {pages} {13628} (\bibinfo {year} {2016})}\BibitemShut
  {NoStop}%
\bibitem [{\citenamefont {Rossi}\ \emph {et~al.}(2017)\citenamefont {Rossi},
  \citenamefont {Kralj}, \citenamefont {Zippilli}, \citenamefont {Natali},
  \citenamefont {Borrielli}, \citenamefont {Pandraud}, \citenamefont {Serra},
  \citenamefont {Di~Giuseppe},\ and\ \citenamefont {Vitali}}]{Vitali2017FB}%
  \BibitemOpen
  \bibfield  {author} {\bibinfo {author} {\bibfnamefont {Massimiliano}\
  \bibnamefont {Rossi}}, \bibinfo {author} {\bibfnamefont {Nenad}\ \bibnamefont
  {Kralj}}, \bibinfo {author} {\bibfnamefont {Stefano}\ \bibnamefont
  {Zippilli}}, \bibinfo {author} {\bibfnamefont {Riccardo}\ \bibnamefont
  {Natali}}, \bibinfo {author} {\bibfnamefont {Antonio}\ \bibnamefont
  {Borrielli}}, \bibinfo {author} {\bibfnamefont {Gregory}\ \bibnamefont
  {Pandraud}}, \bibinfo {author} {\bibfnamefont {Enrico}\ \bibnamefont
  {Serra}}, \bibinfo {author} {\bibfnamefont {Giovanni}\ \bibnamefont
  {Di~Giuseppe}}, \ and\ \bibinfo {author} {\bibfnamefont {David}\ \bibnamefont
  {Vitali}},\ }\bibfield  {title} {\enquote {\bibinfo {title} {Enhancing
  sideband cooling by feedback-controlled light},}\ }\href@noop {} {\bibfield
  {journal} {\bibinfo  {journal} {Phys. Rev. Lett.}\ }\textbf {\bibinfo
  {volume} {119}},\ \bibinfo {pages} {123603} (\bibinfo {year}
  {2017})}\BibitemShut {NoStop}%
\bibitem [{\citenamefont {Sudhir}\ \emph {et~al.}(2017)\citenamefont {Sudhir},
  \citenamefont {Wilson}, \citenamefont {Schilling}, \citenamefont {Schütz},
  \citenamefont {Fedorov},\ and\ \citenamefont
  {Kippenberg}}]{Kippenberg2017SquA}%
  \BibitemOpen
  \bibfield  {author} {\bibinfo {author} {\bibfnamefont {V.}~\bibnamefont
  {Sudhir}}, \bibinfo {author} {\bibfnamefont {D.~J.}\ \bibnamefont {Wilson}},
  \bibinfo {author} {\bibfnamefont {R.}~\bibnamefont {Schilling}}, \bibinfo
  {author} {\bibfnamefont {H.}~\bibnamefont {Schütz}}, \bibinfo {author}
  {\bibfnamefont {S.~A.}\ \bibnamefont {Fedorov}}, \ and\ \bibinfo {author}
  {\bibfnamefont {T.~J.}\ \bibnamefont {Kippenberg}},\ }\bibfield  {title}
  {\enquote {\bibinfo {title} {{Appearance and disappearance of quantum
  correlations in measurement-based feedback control of a mechanical
  oscillator}},}\ }\href@noop {} {\bibfield  {journal} {\bibinfo  {journal}
  {Physical Review X}\ }\textbf {\bibinfo {volume} {7}},\ \bibinfo {pages}
  {011001} (\bibinfo {year} {2017})}\BibitemShut {NoStop}%
\bibitem [{\citenamefont {Christoph}\ \emph {et~al.}(2018)\citenamefont
  {Christoph}, \citenamefont {Wagner}, \citenamefont {Zhong}, \citenamefont
  {Wiesendanger}, \citenamefont {Sengstock}, \citenamefont {Schwarz},\ and\
  \citenamefont {Becker}}]{Becker2018symph}%
  \BibitemOpen
  \bibfield  {author} {\bibinfo {author} {\bibfnamefont {Philipp}\ \bibnamefont
  {Christoph}}, \bibinfo {author} {\bibfnamefont {Tobias}\ \bibnamefont
  {Wagner}}, \bibinfo {author} {\bibfnamefont {Hai}\ \bibnamefont {Zhong}},
  \bibinfo {author} {\bibfnamefont {Roland}\ \bibnamefont {Wiesendanger}},
  \bibinfo {author} {\bibfnamefont {Klaus}\ \bibnamefont {Sengstock}}, \bibinfo
  {author} {\bibfnamefont {Alexander}\ \bibnamefont {Schwarz}}, \ and\ \bibinfo
  {author} {\bibfnamefont {Christoph}\ \bibnamefont {Becker}},\ }\bibfield
  {title} {\enquote {\bibinfo {title} {Combined feedback and sympathetic
  cooling of a mechanical oscillator coupled to ultracold atoms},}\ }\href@noop
  {} {\bibfield  {journal} {\bibinfo  {journal} {New Journal of Physics}\
  }\textbf {\bibinfo {volume} {20}},\ \bibinfo {pages} {093020} (\bibinfo
  {year} {2018})}\BibitemShut {NoStop}%
\bibitem [{\citenamefont {Guo}\ \emph {et~al.}(2019)\citenamefont {Guo},
  \citenamefont {Norte},\ and\ \citenamefont
  {Gr\"oblacher}}]{Groblacher2019FB}%
  \BibitemOpen
  \bibfield  {author} {\bibinfo {author} {\bibfnamefont {Jingkun}\ \bibnamefont
  {Guo}}, \bibinfo {author} {\bibfnamefont {Richard}\ \bibnamefont {Norte}}, \
  and\ \bibinfo {author} {\bibfnamefont {Simon}\ \bibnamefont {Gr\"oblacher}},\
  }\bibfield  {title} {\enquote {\bibinfo {title} {Feedback cooling of a room
  temperature mechanical oscillator close to its motional ground state},}\
  }\href@noop {} {\bibfield  {journal} {\bibinfo  {journal} {Phys. Rev. Lett.}\
  }\textbf {\bibinfo {volume} {123}},\ \bibinfo {pages} {223602} (\bibinfo
  {year} {2019})}\BibitemShut {NoStop}%
\bibitem [{\citenamefont {Kumar}\ \emph {et~al.}(2022)\citenamefont {Kumar},
  \citenamefont {Nätkinniemi}, \citenamefont {Lyyra},\ and\ \citenamefont
  {Muhonen}}]{Muhonen2022FB}%
  \BibitemOpen
  \bibfield  {author} {\bibinfo {author} {\bibfnamefont {Arvind~Shankar}\
  \bibnamefont {Kumar}}, \bibinfo {author} {\bibfnamefont {Joonas}\
  \bibnamefont {Nätkinniemi}}, \bibinfo {author} {\bibfnamefont {Henri}\
  \bibnamefont {Lyyra}}, \ and\ \bibinfo {author} {\bibfnamefont {Juha~T.}\
  \bibnamefont {Muhonen}},\ }\bibfield  {title} {\enquote {\bibinfo {title}
  {{Single-laser feedback cooling of optomechanical resonators}},}\ }\href@noop
  {} {\bibfield  {journal} {\bibinfo  {journal} {arXiv:2209.06029}\ } (\bibinfo
  {year} {2022})}\BibitemShut {NoStop}%
\bibitem [{\citenamefont {Rossi}\ \emph {et~al.}(2018)\citenamefont {Rossi},
  \citenamefont {Mason}, \citenamefont {Chen}, \citenamefont {Tsaturyan},\ and\
  \citenamefont {Schliesser}}]{Schliesser2018FB}%
  \BibitemOpen
  \bibfield  {author} {\bibinfo {author} {\bibfnamefont {Massimiliano}\
  \bibnamefont {Rossi}}, \bibinfo {author} {\bibfnamefont {David}\ \bibnamefont
  {Mason}}, \bibinfo {author} {\bibfnamefont {Junxin}\ \bibnamefont {Chen}},
  \bibinfo {author} {\bibfnamefont {Yeghishe}\ \bibnamefont {Tsaturyan}}, \
  and\ \bibinfo {author} {\bibfnamefont {Albert}\ \bibnamefont {Schliesser}},\
  }\bibfield  {title} {\enquote {\bibinfo {title} {Measurement-based quantum
  control of mechanical motion},}\ }\href@noop {} {\bibfield  {journal}
  {\bibinfo  {journal} {Nature}\ }\textbf {\bibinfo {volume} {563}},\ \bibinfo
  {pages} {53--58} (\bibinfo {year} {2018})}\BibitemShut {NoStop}%
\bibitem [{\citenamefont {Abbott}\ \emph {et~al.}(2009)\citenamefont {Abbott}
  \emph {et~al.}}]{LIGO2009}%
  \BibitemOpen
  \bibfield  {author} {\bibinfo {author} {\bibfnamefont {B}~\bibnamefont
  {Abbott}} \emph {et~al.},\ }\bibfield  {title} {\enquote {\bibinfo {title}
  {Observation of a kilogram-scale oscillator near its quantum ground state},}\
  }\href@noop {} {\bibfield  {journal} {\bibinfo  {journal} {New Journal of
  Physics}\ }\textbf {\bibinfo {volume} {11}},\ \bibinfo {pages} {073032}
  (\bibinfo {year} {2009})}\BibitemShut {NoStop}%
\bibitem [{\citenamefont {Whittle}\ \emph {et~al.}(2021)\citenamefont {Whittle}
  \emph {et~al.}}]{LIGO2021}%
  \BibitemOpen
  \bibfield  {author} {\bibinfo {author} {\bibfnamefont {C.}~\bibnamefont
  {Whittle}} \emph {et~al.},\ }\bibfield  {title} {\enquote {\bibinfo {title}
  {Approaching the motional ground state of a 10-kg object},}\ }\href@noop {}
  {\bibfield  {journal} {\bibinfo  {journal} {Science}\ }\textbf {\bibinfo
  {volume} {372}},\ \bibinfo {pages} {1333--1336} (\bibinfo {year}
  {2021})}\BibitemShut {NoStop}%
\bibitem [{\citenamefont {Gieseler}\ \emph {et~al.}(2012)\citenamefont
  {Gieseler}, \citenamefont {Deutsch}, \citenamefont {Quidant},\ and\
  \citenamefont {Novotny}}]{Novotny2012FB}%
  \BibitemOpen
  \bibfield  {author} {\bibinfo {author} {\bibfnamefont {Jan}\ \bibnamefont
  {Gieseler}}, \bibinfo {author} {\bibfnamefont {Bradley}\ \bibnamefont
  {Deutsch}}, \bibinfo {author} {\bibfnamefont {Romain}\ \bibnamefont
  {Quidant}}, \ and\ \bibinfo {author} {\bibfnamefont {Lukas}\ \bibnamefont
  {Novotny}},\ }\bibfield  {title} {\enquote {\bibinfo {title} {Subkelvin
  parametric feedback cooling of a laser-trapped nanoparticle},}\ }\href@noop
  {} {\bibfield  {journal} {\bibinfo  {journal} {Phys. Rev. Lett.}\ }\textbf
  {\bibinfo {volume} {109}},\ \bibinfo {pages} {103603} (\bibinfo {year}
  {2012})}\BibitemShut {NoStop}%
\bibitem [{\citenamefont {Vovrosh}\ \emph {et~al.}(2017)\citenamefont
  {Vovrosh}, \citenamefont {Rashid}, \citenamefont {Hempston}, \citenamefont
  {Bateman}, \citenamefont {Paternostro},\ and\ \citenamefont
  {Ulbricht}}]{Paternostro2017FB}%
  \BibitemOpen
  \bibfield  {author} {\bibinfo {author} {\bibfnamefont {Jamie}\ \bibnamefont
  {Vovrosh}}, \bibinfo {author} {\bibfnamefont {Muddassar}\ \bibnamefont
  {Rashid}}, \bibinfo {author} {\bibfnamefont {David}\ \bibnamefont
  {Hempston}}, \bibinfo {author} {\bibfnamefont {James}\ \bibnamefont
  {Bateman}}, \bibinfo {author} {\bibfnamefont {Mauro}\ \bibnamefont
  {Paternostro}}, \ and\ \bibinfo {author} {\bibfnamefont {Hendrik}\
  \bibnamefont {Ulbricht}},\ }\bibfield  {title} {\enquote {\bibinfo {title}
  {Parametric feedback cooling of levitated optomechanics in a parabolic mirror
  trap},}\ }\href@noop {} {\bibfield  {journal} {\bibinfo  {journal} {J. Opt.
  Soc. Am. B}\ }\textbf {\bibinfo {volume} {34}},\ \bibinfo {pages}
  {1421--1428} (\bibinfo {year} {2017})}\BibitemShut {NoStop}%
\bibitem [{\citenamefont {Li}\ \emph {et~al.}(2011)\citenamefont {Li},
  \citenamefont {Kheifets},\ and\ \citenamefont {Raizen}}]{Raizen2011Levit}%
  \BibitemOpen
  \bibfield  {author} {\bibinfo {author} {\bibfnamefont {Tongcang}\
  \bibnamefont {Li}}, \bibinfo {author} {\bibfnamefont {Simon}\ \bibnamefont
  {Kheifets}}, \ and\ \bibinfo {author} {\bibfnamefont {Mark~G.}\ \bibnamefont
  {Raizen}},\ }\bibfield  {title} {\enquote {\bibinfo {title} {Millikelvin
  cooling of an optically trapped microsphere in vacuum},}\ }\href@noop {}
  {\bibfield  {journal} {\bibinfo  {journal} {Nature Physics}\ }\textbf
  {\bibinfo {volume} {7}},\ \bibinfo {pages} {527--530} (\bibinfo {year}
  {2011})}\BibitemShut {NoStop}%
\bibitem [{\citenamefont {Conangla}\ \emph {et~al.}(2019)\citenamefont
  {Conangla}, \citenamefont {Ricci}, \citenamefont {Cuairan}, \citenamefont
  {Schell}, \citenamefont {Meyer},\ and\ \citenamefont
  {Quidant}}]{Quidant2019FB}%
  \BibitemOpen
  \bibfield  {author} {\bibinfo {author} {\bibfnamefont {Gerard~P.}\
  \bibnamefont {Conangla}}, \bibinfo {author} {\bibfnamefont {Francesco}\
  \bibnamefont {Ricci}}, \bibinfo {author} {\bibfnamefont {Marc~T.}\
  \bibnamefont {Cuairan}}, \bibinfo {author} {\bibfnamefont {Andreas~W.}\
  \bibnamefont {Schell}}, \bibinfo {author} {\bibfnamefont {Nadine}\
  \bibnamefont {Meyer}}, \ and\ \bibinfo {author} {\bibfnamefont {Romain}\
  \bibnamefont {Quidant}},\ }\bibfield  {title} {\enquote {\bibinfo {title}
  {Optimal feedback cooling of a charged levitated nanoparticle with adaptive
  control},}\ }\href@noop {} {\bibfield  {journal} {\bibinfo  {journal} {Phys.
  Rev. Lett.}\ }\textbf {\bibinfo {volume} {122}},\ \bibinfo {pages} {223602}
  (\bibinfo {year} {2019})}\BibitemShut {NoStop}%
\bibitem [{\citenamefont {Iwasaki}\ \emph {et~al.}(2019)\citenamefont
  {Iwasaki}, \citenamefont {Yotsuya}, \citenamefont {Naruki}, \citenamefont
  {Matsuda}, \citenamefont {Yoneda},\ and\ \citenamefont
  {Aikawa}}]{Iwasaki2019FB}%
  \BibitemOpen
  \bibfield  {author} {\bibinfo {author} {\bibfnamefont {M.}~\bibnamefont
  {Iwasaki}}, \bibinfo {author} {\bibfnamefont {T.}~\bibnamefont {Yotsuya}},
  \bibinfo {author} {\bibfnamefont {T.}~\bibnamefont {Naruki}}, \bibinfo
  {author} {\bibfnamefont {Y.}~\bibnamefont {Matsuda}}, \bibinfo {author}
  {\bibfnamefont {M.}~\bibnamefont {Yoneda}}, \ and\ \bibinfo {author}
  {\bibfnamefont {K.}~\bibnamefont {Aikawa}},\ }\bibfield  {title} {\enquote
  {\bibinfo {title} {Electric feedback cooling of single charged nanoparticles
  in an optical trap},}\ }\href@noop {} {\bibfield  {journal} {\bibinfo
  {journal} {Phys. Rev. A}\ }\textbf {\bibinfo {volume} {99}},\ \bibinfo
  {pages} {051401} (\bibinfo {year} {2019})}\BibitemShut {NoStop}%
\bibitem [{\citenamefont {Deli{\'c}}\ \emph {et~al.}(2020)\citenamefont
  {Deli{\'c}}, \citenamefont {Reisenbauer}, \citenamefont {Dare}, \citenamefont
  {Grass}, \citenamefont {Vuleti{\'c}}, \citenamefont {Kiesel},\ and\
  \citenamefont {Aspelmeyer}}]{Aspelmeyer2020Levit}%
  \BibitemOpen
  \bibfield  {author} {\bibinfo {author} {\bibfnamefont {Uro{\v s}}\
  \bibnamefont {Deli{\'c}}}, \bibinfo {author} {\bibfnamefont {Manuel}\
  \bibnamefont {Reisenbauer}}, \bibinfo {author} {\bibfnamefont {Kahan}\
  \bibnamefont {Dare}}, \bibinfo {author} {\bibfnamefont {David}\ \bibnamefont
  {Grass}}, \bibinfo {author} {\bibfnamefont {Vladan}\ \bibnamefont
  {Vuleti{\'c}}}, \bibinfo {author} {\bibfnamefont {Nikolai}\ \bibnamefont
  {Kiesel}}, \ and\ \bibinfo {author} {\bibfnamefont {Markus}\ \bibnamefont
  {Aspelmeyer}},\ }\bibfield  {title} {\enquote {\bibinfo {title} {Cooling of a
  levitated nanoparticle to the motional quantum ground state},}\ }\href@noop
  {} {\bibfield  {journal} {\bibinfo  {journal} {Science}\ }\textbf {\bibinfo
  {volume} {367}},\ \bibinfo {pages} {892--895} (\bibinfo {year}
  {2020})}\BibitemShut {NoStop}%
\bibitem [{\citenamefont {Tebbenjohanns}\ \emph {et~al.}(2021)\citenamefont
  {Tebbenjohanns}, \citenamefont {Mattana}, \citenamefont {Rossi},
  \citenamefont {Frimmer},\ and\ \citenamefont {Novotny}}]{Novotny2021FB}%
  \BibitemOpen
  \bibfield  {author} {\bibinfo {author} {\bibfnamefont {Felix}\ \bibnamefont
  {Tebbenjohanns}}, \bibinfo {author} {\bibfnamefont {M.~Luisa}\ \bibnamefont
  {Mattana}}, \bibinfo {author} {\bibfnamefont {Massimiliano}\ \bibnamefont
  {Rossi}}, \bibinfo {author} {\bibfnamefont {Martin}\ \bibnamefont {Frimmer}},
  \ and\ \bibinfo {author} {\bibfnamefont {Lukas}\ \bibnamefont {Novotny}},\
  }\bibfield  {title} {\enquote {\bibinfo {title} {Quantum control of a
  nanoparticle optically levitated in cryogenic free space},}\ }\href@noop {}
  {\bibfield  {journal} {\bibinfo  {journal} {Nature}\ }\textbf {\bibinfo
  {volume} {595}},\ \bibinfo {pages} {378--382} (\bibinfo {year}
  {2021})}\BibitemShut {NoStop}%
\bibitem [{\citenamefont {Regal}\ \emph {et~al.}(2008)\citenamefont {Regal},
  \citenamefont {Teufel},\ and\ \citenamefont {Lehnert}}]{Lehnert2008Nph}%
  \BibitemOpen
  \bibfield  {author} {\bibinfo {author} {\bibfnamefont {C.~A.}\ \bibnamefont
  {Regal}}, \bibinfo {author} {\bibfnamefont {J~D}\ \bibnamefont {Teufel}}, \
  and\ \bibinfo {author} {\bibfnamefont {K.~W.}\ \bibnamefont {Lehnert}},\
  }\bibfield  {title} {\enquote {\bibinfo {title} {{Measuring nanomechanical
  motion with a microwave cavity interferometer}},}\ }\href@noop {} {\bibfield
  {journal} {\bibinfo  {journal} {Nature Physics}\ }\textbf {\bibinfo {volume}
  {4}},\ \bibinfo {pages} {555--560} (\bibinfo {year} {2008})}\BibitemShut
  {NoStop}%
\bibitem [{\citenamefont {Teufel}\ \emph {et~al.}(2011)\citenamefont {Teufel},
  \citenamefont {Donner}, \citenamefont {Li}, \citenamefont {Harlow},
  \citenamefont {Allman}, \citenamefont {Cicak}, \citenamefont {Sirois},
  \citenamefont {Whittaker}, \citenamefont {Lehnert},\ and\ \citenamefont
  {Simmonds}}]{Teufel2011b}%
  \BibitemOpen
  \bibfield  {author} {\bibinfo {author} {\bibfnamefont {J.~D.}\ \bibnamefont
  {Teufel}}, \bibinfo {author} {\bibfnamefont {T.}~\bibnamefont {Donner}},
  \bibinfo {author} {\bibfnamefont {Dale}\ \bibnamefont {Li}}, \bibinfo
  {author} {\bibfnamefont {J.~W.}\ \bibnamefont {Harlow}}, \bibinfo {author}
  {\bibfnamefont {M.~S.}\ \bibnamefont {Allman}}, \bibinfo {author}
  {\bibfnamefont {K.}~\bibnamefont {Cicak}}, \bibinfo {author} {\bibfnamefont
  {A.~J.}\ \bibnamefont {Sirois}}, \bibinfo {author} {\bibfnamefont {J.~D.}\
  \bibnamefont {Whittaker}}, \bibinfo {author} {\bibfnamefont {K.~W.}\
  \bibnamefont {Lehnert}}, \ and\ \bibinfo {author} {\bibfnamefont {R.~W.}\
  \bibnamefont {Simmonds}},\ }\bibfield  {title} {\enquote {\bibinfo {title}
  {{Sideband cooling of micromechanical motion to the quantum ground state}},}\
  }\href@noop {} {\bibfield  {journal} {\bibinfo  {journal} {Nature}\ }\textbf
  {\bibinfo {volume} {475}},\ \bibinfo {pages} {359--363} (\bibinfo {year}
  {2011})}\BibitemShut {NoStop}%
\bibitem [{\citenamefont {Mancini}\ \emph {et~al.}(1998)\citenamefont
  {Mancini}, \citenamefont {Vitali},\ and\ \citenamefont
  {Tombesi}}]{Tombesi1998FB}%
  \BibitemOpen
  \bibfield  {author} {\bibinfo {author} {\bibfnamefont {Stefano}\ \bibnamefont
  {Mancini}}, \bibinfo {author} {\bibfnamefont {David}\ \bibnamefont {Vitali}},
  \ and\ \bibinfo {author} {\bibfnamefont {Paolo}\ \bibnamefont {Tombesi}},\
  }\bibfield  {title} {\enquote {\bibinfo {title} {Optomechanical cooling of a
  macroscopic oscillator by homodyne feedback},}\ }\href@noop {} {\bibfield
  {journal} {\bibinfo  {journal} {Phys. Rev. Lett.}\ }\textbf {\bibinfo
  {volume} {80}},\ \bibinfo {pages} {688--691} (\bibinfo {year}
  {1998})}\BibitemShut {NoStop}%
\bibitem [{\citenamefont {Vitali}\ \emph {et~al.}(2002)\citenamefont {Vitali},
  \citenamefont {Mancini}, \citenamefont {Ribichini},\ and\ \citenamefont
  {Tombesi}}]{Vitali2002FB}%
  \BibitemOpen
  \bibfield  {author} {\bibinfo {author} {\bibfnamefont {David}\ \bibnamefont
  {Vitali}}, \bibinfo {author} {\bibfnamefont {Stefano}\ \bibnamefont
  {Mancini}}, \bibinfo {author} {\bibfnamefont {Luciano}\ \bibnamefont
  {Ribichini}}, \ and\ \bibinfo {author} {\bibfnamefont {Paolo}\ \bibnamefont
  {Tombesi}},\ }\bibfield  {title} {\enquote {\bibinfo {title} {Mirror
  quiescence and high-sensitivity position measurements with feedback},}\
  }\href@noop {} {\bibfield  {journal} {\bibinfo  {journal} {Phys. Rev. A}\
  }\textbf {\bibinfo {volume} {65}},\ \bibinfo {pages} {063803} (\bibinfo
  {year} {2002})}\BibitemShut {NoStop}%
\bibitem [{\citenamefont {Hopkins}\ \emph {et~al.}(2003)\citenamefont
  {Hopkins}, \citenamefont {Jacobs}, \citenamefont {Habib},\ and\ \citenamefont
  {Schwab}}]{Schwab2003FB}%
  \BibitemOpen
  \bibfield  {author} {\bibinfo {author} {\bibfnamefont {Asa}\ \bibnamefont
  {Hopkins}}, \bibinfo {author} {\bibfnamefont {Kurt}\ \bibnamefont {Jacobs}},
  \bibinfo {author} {\bibfnamefont {Salman}\ \bibnamefont {Habib}}, \ and\
  \bibinfo {author} {\bibfnamefont {Keith}\ \bibnamefont {Schwab}},\ }\bibfield
   {title} {\enquote {\bibinfo {title} {Feedback cooling of a nanomechanical
  resonator},}\ }\href@noop {} {\bibfield  {journal} {\bibinfo  {journal}
  {Phys. Rev. B}\ }\textbf {\bibinfo {volume} {68}},\ \bibinfo {pages} {235328}
  (\bibinfo {year} {2003})}\BibitemShut {NoStop}%
\bibitem [{\citenamefont {Genes}\ \emph {et~al.}(2008)\citenamefont {Genes},
  \citenamefont {Vitali}, \citenamefont {Tombesi},\ and\ \citenamefont
  {Aspelmeyer}}]{Aspelmeyer2008Cool}%
  \BibitemOpen
  \bibfield  {author} {\bibinfo {author} {\bibfnamefont {C.}~\bibnamefont
  {Genes}}, \bibinfo {author} {\bibfnamefont {D.}~\bibnamefont {Vitali}},
  \bibinfo {author} {\bibfnamefont {P.}~\bibnamefont {Tombesi}}, \ and\
  \bibinfo {author} {\bibfnamefont {M.}~\bibnamefont {Aspelmeyer}},\ }\bibfield
   {title} {\enquote {\bibinfo {title} {{Ground-state cooling of a
  micromechanical oscillator: Comparing cold damping and cavity-assisted
  cooling schemes}},}\ }\href@noop {} {\bibfield  {journal} {\bibinfo
  {journal} {Physical Review A}\ }\textbf {\bibinfo {volume} {77}},\ \bibinfo
  {pages} {033804} (\bibinfo {year} {2008})}\BibitemShut {NoStop}%
\bibitem [{\citenamefont {Zhang}\ \emph {et~al.}(2017)\citenamefont {Zhang},
  \citenamefont {xi~Liu}, \citenamefont {Wu}, \citenamefont {Jacobs},\ and\
  \citenamefont {Nori}}]{Nori2017FBreview}%
  \BibitemOpen
  \bibfield  {author} {\bibinfo {author} {\bibfnamefont {Jing}\ \bibnamefont
  {Zhang}}, \bibinfo {author} {\bibfnamefont {Yu}~\bibnamefont {xi~Liu}},
  \bibinfo {author} {\bibfnamefont {Re-Bing}\ \bibnamefont {Wu}}, \bibinfo
  {author} {\bibfnamefont {Kurt}\ \bibnamefont {Jacobs}}, \ and\ \bibinfo
  {author} {\bibfnamefont {Franco}\ \bibnamefont {Nori}},\ }\bibfield  {title}
  {\enquote {\bibinfo {title} {Quantum feedback: Theory, experiments, and
  applications},}\ }\href@noop {} {\bibfield  {journal} {\bibinfo  {journal}
  {Physics Reports}\ }\textbf {\bibinfo {volume} {679}},\ \bibinfo {pages}
  {1--60} (\bibinfo {year} {2017})}\BibitemShut {NoStop}%
\bibitem [{\citenamefont {Sommer}\ and\ \citenamefont
  {Genes}(2019)}]{Genes2019FB}%
  \BibitemOpen
  \bibfield  {author} {\bibinfo {author} {\bibfnamefont {Christian}\
  \bibnamefont {Sommer}}\ and\ \bibinfo {author} {\bibfnamefont {Claudiu}\
  \bibnamefont {Genes}},\ }\bibfield  {title} {\enquote {\bibinfo {title}
  {Partial optomechanical refrigeration via multimode cold-damping feedback},}\
  }\href@noop {} {\bibfield  {journal} {\bibinfo  {journal} {Phys. Rev. Lett.}\
  }\textbf {\bibinfo {volume} {123}},\ \bibinfo {pages} {203605} (\bibinfo
  {year} {2019})}\BibitemShut {NoStop}%
\bibitem [{\citenamefont {Woolley}\ and\ \citenamefont
  {Clerk}(2013)}]{WoolleyBAE}%
  \BibitemOpen
  \bibfield  {author} {\bibinfo {author} {\bibfnamefont {M.~J.}\ \bibnamefont
  {Woolley}}\ and\ \bibinfo {author} {\bibfnamefont {A.~A.}\ \bibnamefont
  {Clerk}},\ }\bibfield  {title} {\enquote {\bibinfo {title} {Two-mode
  back-action-evading measurements in cavity optomechanics},}\ }\href@noop {}
  {\bibfield  {journal} {\bibinfo  {journal} {Phys. Rev. A}\ }\textbf {\bibinfo
  {volume} {87}},\ \bibinfo {pages} {063846} (\bibinfo {year}
  {2013})}\BibitemShut {NoStop}%
\bibitem [{\citenamefont {Wollman}\ \emph {et~al.}(2015)\citenamefont
  {Wollman}, \citenamefont {Lei}, \citenamefont {Weinstein}, \citenamefont
  {Suh}, \citenamefont {Kronwald}, \citenamefont {Marquardt}, \citenamefont
  {Clerk},\ and\ \citenamefont {Schwab}}]{SchwabSqueeze}%
  \BibitemOpen
  \bibfield  {author} {\bibinfo {author} {\bibfnamefont {E.~E.}\ \bibnamefont
  {Wollman}}, \bibinfo {author} {\bibfnamefont {C.~U.}\ \bibnamefont {Lei}},
  \bibinfo {author} {\bibfnamefont {A.~J.}\ \bibnamefont {Weinstein}}, \bibinfo
  {author} {\bibfnamefont {J.}~\bibnamefont {Suh}}, \bibinfo {author}
  {\bibfnamefont {A.}~\bibnamefont {Kronwald}}, \bibinfo {author}
  {\bibfnamefont {F.}~\bibnamefont {Marquardt}}, \bibinfo {author}
  {\bibfnamefont {A.~A.}\ \bibnamefont {Clerk}}, \ and\ \bibinfo {author}
  {\bibfnamefont {K.~C.}\ \bibnamefont {Schwab}},\ }\bibfield  {title}
  {\enquote {\bibinfo {title} {Quantum squeezing of motion in a mechanical
  resonator},}\ }\href@noop {} {\bibfield  {journal} {\bibinfo  {journal}
  {Science}\ }\textbf {\bibinfo {volume} {349}},\ \bibinfo {pages} {952--955}
  (\bibinfo {year} {2015})}\BibitemShut {NoStop}%
\bibitem [{\citenamefont {Lecocq}\ \emph
  {et~al.}(2015{\natexlab{a}})\citenamefont {Lecocq}, \citenamefont {Clark},
  \citenamefont {Simmonds}, \citenamefont {Aumentado},\ and\ \citenamefont
  {Teufel}}]{TeufelSqueeze}%
  \BibitemOpen
  \bibfield  {author} {\bibinfo {author} {\bibfnamefont {F.}~\bibnamefont
  {Lecocq}}, \bibinfo {author} {\bibfnamefont {J.~B.}\ \bibnamefont {Clark}},
  \bibinfo {author} {\bibfnamefont {R.~W.}\ \bibnamefont {Simmonds}}, \bibinfo
  {author} {\bibfnamefont {J.}~\bibnamefont {Aumentado}}, \ and\ \bibinfo
  {author} {\bibfnamefont {J.~D.}\ \bibnamefont {Teufel}},\ }\bibfield  {title}
  {\enquote {\bibinfo {title} {Quantum nondemolition measurement of a
  nonclassical state of a massive object},}\ }\href@noop {} {\bibfield
  {journal} {\bibinfo  {journal} {Phys. Rev. X}\ }\textbf {\bibinfo {volume}
  {5}},\ \bibinfo {pages} {041037} (\bibinfo {year}
  {2015}{\natexlab{a}})}\BibitemShut {NoStop}%
\bibitem [{\citenamefont {Pirkkalainen}\ \emph {et~al.}(2015)\citenamefont
  {Pirkkalainen}, \citenamefont {Damsk\"agg}, \citenamefont {Brandt},
  \citenamefont {Massel},\ and\ \citenamefont {Sillanp\"a\"a}}]{Squeeze}%
  \BibitemOpen
  \bibfield  {author} {\bibinfo {author} {\bibfnamefont {J.-M.}\ \bibnamefont
  {Pirkkalainen}}, \bibinfo {author} {\bibfnamefont {E.}~\bibnamefont
  {Damsk\"agg}}, \bibinfo {author} {\bibfnamefont {M.}~\bibnamefont {Brandt}},
  \bibinfo {author} {\bibfnamefont {F.}~\bibnamefont {Massel}}, \ and\ \bibinfo
  {author} {\bibfnamefont {M.~A.}\ \bibnamefont {Sillanp\"a\"a}},\ }\bibfield
  {title} {\enquote {\bibinfo {title} {Squeezing of quantum noise of motion in
  a micromechanical resonator},}\ }\href@noop {} {\bibfield  {journal}
  {\bibinfo  {journal} {Phys.~Rev.~Lett.}\ }\textbf {\bibinfo {volume} {115}},\
  \bibinfo {pages} {243601} (\bibinfo {year} {2015})}\BibitemShut {NoStop}%
\bibitem [{\citenamefont {Ockeloen-Korppi}\ \emph {et~al.}(2018)\citenamefont
  {Ockeloen-Korppi}, \citenamefont {Damsk{\"a}gg}, \citenamefont
  {Pirkkalainen}, \citenamefont {Asjad}, \citenamefont {Clerk}, \citenamefont
  {Massel}, \citenamefont {Woolley},\ and\ \citenamefont
  {Sillanp{\"a}{\"a}}}]{Entanglement}%
  \BibitemOpen
  \bibfield  {author} {\bibinfo {author} {\bibfnamefont {C.~F.}\ \bibnamefont
  {Ockeloen-Korppi}}, \bibinfo {author} {\bibfnamefont {E.}~\bibnamefont
  {Damsk{\"a}gg}}, \bibinfo {author} {\bibfnamefont {J.~M.}\ \bibnamefont
  {Pirkkalainen}}, \bibinfo {author} {\bibfnamefont {M.}~\bibnamefont {Asjad}},
  \bibinfo {author} {\bibfnamefont {A.~A.}\ \bibnamefont {Clerk}}, \bibinfo
  {author} {\bibfnamefont {F.}~\bibnamefont {Massel}}, \bibinfo {author}
  {\bibfnamefont {M.~J.}\ \bibnamefont {Woolley}}, \ and\ \bibinfo {author}
  {\bibfnamefont {M.~A.}\ \bibnamefont {Sillanp{\"a}{\"a}}},\ }\bibfield
  {title} {\enquote {\bibinfo {title} {Stabilized entanglement of massive
  mechanical oscillators},}\ }\href@noop {} {\bibfield  {journal} {\bibinfo
  {journal} {Nature}\ }\textbf {\bibinfo {volume} {556}},\ \bibinfo {pages}
  {478--482} (\bibinfo {year} {2018})}\BibitemShut {NoStop}%
\bibitem [{\citenamefont {Zhou}\ \emph {et~al.}(2019)\citenamefont {Zhou},
  \citenamefont {Cattiaux}, \citenamefont {Gazizulin}, \citenamefont {Luck},
  \citenamefont {Maillet}, \citenamefont {Crozes}, \citenamefont {Motte},
  \citenamefont {Bourgeois}, \citenamefont {Fefferman},\ and\ \citenamefont
  {Collin}}]{Collin2019demag}%
  \BibitemOpen
  \bibfield  {author} {\bibinfo {author} {\bibfnamefont {X.}~\bibnamefont
  {Zhou}}, \bibinfo {author} {\bibfnamefont {D.}~\bibnamefont {Cattiaux}},
  \bibinfo {author} {\bibfnamefont {R.~R.}\ \bibnamefont {Gazizulin}}, \bibinfo
  {author} {\bibfnamefont {A.}~\bibnamefont {Luck}}, \bibinfo {author}
  {\bibfnamefont {O.}~\bibnamefont {Maillet}}, \bibinfo {author} {\bibfnamefont
  {T.}~\bibnamefont {Crozes}}, \bibinfo {author} {\bibfnamefont {J.-F.}\
  \bibnamefont {Motte}}, \bibinfo {author} {\bibfnamefont {O.}~\bibnamefont
  {Bourgeois}}, \bibinfo {author} {\bibfnamefont {A.}~\bibnamefont
  {Fefferman}}, \ and\ \bibinfo {author} {\bibfnamefont {E.}~\bibnamefont
  {Collin}},\ }\bibfield  {title} {\enquote {\bibinfo {title} {On-chip
  thermometry for microwave optomechanics implemented in a nuclear
  demagnetization cryostat},}\ }\href@noop {} {\bibfield  {journal} {\bibinfo
  {journal} {Phys. Rev. Applied}\ }\textbf {\bibinfo {volume} {12}},\ \bibinfo
  {pages} {044066} (\bibinfo {year} {2019})}\BibitemShut {NoStop}%
\bibitem [{\citenamefont {Weinstein}\ \emph {et~al.}(2014)\citenamefont
  {Weinstein}, \citenamefont {Lei}, \citenamefont {Wollman}, \citenamefont
  {Suh}, \citenamefont {Metelmann}, \citenamefont {Clerk},\ and\ \citenamefont
  {Schwab}}]{Schwab2014Asymm}%
  \BibitemOpen
  \bibfield  {author} {\bibinfo {author} {\bibfnamefont {A.~J.}\ \bibnamefont
  {Weinstein}}, \bibinfo {author} {\bibfnamefont {C.~U.}\ \bibnamefont {Lei}},
  \bibinfo {author} {\bibfnamefont {E.~E.}\ \bibnamefont {Wollman}}, \bibinfo
  {author} {\bibfnamefont {J.}~\bibnamefont {Suh}}, \bibinfo {author}
  {\bibfnamefont {A.}~\bibnamefont {Metelmann}}, \bibinfo {author}
  {\bibfnamefont {A.~A.}\ \bibnamefont {Clerk}}, \ and\ \bibinfo {author}
  {\bibfnamefont {K.~C.}\ \bibnamefont {Schwab}},\ }\bibfield  {title}
  {\enquote {\bibinfo {title} {Observation and interpretation of motional
  sideband asymmetry in a quantum electromechanical device},}\ }\href@noop {}
  {\bibfield  {journal} {\bibinfo  {journal} {Phys. Rev. X}\ }\textbf {\bibinfo
  {volume} {4}},\ \bibinfo {pages} {041003} (\bibinfo {year}
  {2014})}\BibitemShut {NoStop}%
\bibitem [{\citenamefont {Underwood}\ \emph {et~al.}(2015)\citenamefont
  {Underwood}, \citenamefont {Mason}, \citenamefont {Lee}, \citenamefont {Xu},
  \citenamefont {Jiang}, \citenamefont {Shkarin}, \citenamefont {B\o{}rkje},
  \citenamefont {Girvin},\ and\ \citenamefont {Harris}}]{Harris2015asym}%
  \BibitemOpen
  \bibfield  {author} {\bibinfo {author} {\bibfnamefont {M.}~\bibnamefont
  {Underwood}}, \bibinfo {author} {\bibfnamefont {D.}~\bibnamefont {Mason}},
  \bibinfo {author} {\bibfnamefont {D.}~\bibnamefont {Lee}}, \bibinfo {author}
  {\bibfnamefont {H.}~\bibnamefont {Xu}}, \bibinfo {author} {\bibfnamefont
  {L.}~\bibnamefont {Jiang}}, \bibinfo {author} {\bibfnamefont {A.~B.}\
  \bibnamefont {Shkarin}}, \bibinfo {author} {\bibfnamefont {K.}~\bibnamefont
  {B\o{}rkje}}, \bibinfo {author} {\bibfnamefont {S.~M.}\ \bibnamefont
  {Girvin}}, \ and\ \bibinfo {author} {\bibfnamefont {J.~G.~E.}\ \bibnamefont
  {Harris}},\ }\bibfield  {title} {\enquote {\bibinfo {title} {Measurement of
  the motional sidebands of a nanogram-scale oscillator in the quantum
  regime},}\ }\href@noop {} {\bibfield  {journal} {\bibinfo  {journal} {Phys.
  Rev. A}\ }\textbf {\bibinfo {volume} {92}},\ \bibinfo {pages} {061801}
  (\bibinfo {year} {2015})}\BibitemShut {NoStop}%
\bibitem [{\citenamefont {Lecocq}\ \emph
  {et~al.}(2015{\natexlab{b}})\citenamefont {Lecocq}, \citenamefont {Teufel},
  \citenamefont {Aumentado},\ and\ \citenamefont {Simmonds}}]{Simmonds2015qb}%
  \BibitemOpen
  \bibfield  {author} {\bibinfo {author} {\bibfnamefont {F.}~\bibnamefont
  {Lecocq}}, \bibinfo {author} {\bibfnamefont {J.~D.}\ \bibnamefont {Teufel}},
  \bibinfo {author} {\bibfnamefont {J.}~\bibnamefont {Aumentado}}, \ and\
  \bibinfo {author} {\bibfnamefont {R.~W.}\ \bibnamefont {Simmonds}},\
  }\bibfield  {title} {\enquote {\bibinfo {title} {Resolving the vacuum
  fluctuations of an optomechanical system using an artificial atom},}\
  }\href@noop {} {\bibfield  {journal} {\bibinfo  {journal} {Nat. Phys.}\
  }\textbf {\bibinfo {volume} {11}},\ \bibinfo {pages} {635--639} (\bibinfo
  {year} {2015}{\natexlab{b}})}\BibitemShut {NoStop}%
\bibitem [{\citenamefont {Macklin}\ \emph {et~al.}(2015)\citenamefont
  {Macklin}, \citenamefont {O'Brien}, \citenamefont {Hover}, \citenamefont
  {Schwartz}, \citenamefont {Bolkhovsky}, \citenamefont {Zhang}, \citenamefont
  {Oliver},\ and\ \citenamefont {Siddiqi}}]{Siddiqi2015Amp}%
  \BibitemOpen
  \bibfield  {author} {\bibinfo {author} {\bibfnamefont {C.}~\bibnamefont
  {Macklin}}, \bibinfo {author} {\bibfnamefont {K.}~\bibnamefont {O'Brien}},
  \bibinfo {author} {\bibfnamefont {D.}~\bibnamefont {Hover}}, \bibinfo
  {author} {\bibfnamefont {M.~E.}\ \bibnamefont {Schwartz}}, \bibinfo {author}
  {\bibfnamefont {V.}~\bibnamefont {Bolkhovsky}}, \bibinfo {author}
  {\bibfnamefont {X.}~\bibnamefont {Zhang}}, \bibinfo {author} {\bibfnamefont
  {W.~D.}\ \bibnamefont {Oliver}}, \ and\ \bibinfo {author} {\bibfnamefont
  {I.}~\bibnamefont {Siddiqi}},\ }\bibfield  {title} {\enquote {\bibinfo
  {title} {{A near-quantum-limited Josephson traveling-wave parametric
  amplifier}},}\ }\href@noop {} {\bibfield  {journal} {\bibinfo  {journal}
  {Science}\ }\textbf {\bibinfo {volume} {350}},\ \bibinfo {pages} {307--310}
  (\bibinfo {year} {2015})}\BibitemShut {NoStop}%
\bibitem [{\citenamefont {Clerk}\ \emph {et~al.}(2008)\citenamefont {Clerk},
  \citenamefont {Marquardt},\ and\ \citenamefont {Jacobs}}]{Marquardt2008Sq}%
  \BibitemOpen
  \bibfield  {author} {\bibinfo {author} {\bibfnamefont {A.~A.}\ \bibnamefont
  {Clerk}}, \bibinfo {author} {\bibfnamefont {F.}~\bibnamefont {Marquardt}}, \
  and\ \bibinfo {author} {\bibfnamefont {K.}~\bibnamefont {Jacobs}},\
  }\bibfield  {title} {\enquote {\bibinfo {title} {Back-action evasion and
  squeezing of a mechanical resonator using a cavity detector},}\ }\href@noop
  {} {\bibfield  {journal} {\bibinfo  {journal} {New Journal of Physics}\
  }\textbf {\bibinfo {volume} {10}},\ \bibinfo {pages} {095010} (\bibinfo
  {year} {2008})}\BibitemShut {NoStop}%
\end{thebibliography}

%

\appendix

\section{Solving the closed-loop dynamics}
\label{sec:fbnumerics}

In the bad-cavity case there are simple analytical results for the entire system including the output spectrum due to the probe tone (Appendix \ref{sec:badcavityresults}). Also, with arbitrary sideband resolution but restricted to $\Delta = 0$, we can recover expressions for the mechanical occupation, \eref{eq:nmFB}, and the preceding equations. However, with the current parameters we calculate the output spectrum always numerically. We start from \eref{eqmot}, and write the variables' connection to the inputs with coefficients to be determined:
\begin{subequations}
\label{eq:FBcoeff0}
\begin{alignat}{2}
x_c & = X_{cx} x_{\m{c,in}} \,,\\
y_c & = Y_{cy} y_{\m{c,in}}+ Y_x x_{\m{c,in}} + Y_f f_{\m{th}} + Y_n y_{\m{add}} \,,\\
x & = X_f f_{\m{th}} + X_{bax} x_{\m{c,in}} + X_{\m{inj}} y_{\m{c,in}} + X_n y_{\m{add}}  \,, \\
p & = P_f f_{\m{th}} + P_{bax} x_{\m{c,in}} +P_{\m{inj}} y_{\m{c,in}} + P_n y_{\m{add}}  \,.
\end{alignat}
\end{subequations}
As an example, the term with $X_{bax}$ gives the quantum measurement backaction to the position, $X_{\m{inj}}$ describes how the quantum noise in the feedback loop is injected back to the sample, and $X_n = X_{\m{inj}}$ is a similar injection of amplifier noise. To continue with the example, in the case $\Delta = 0$,
\begin{equation}
\begin{split}
X_f & =  \chi_{\m{fb}} \,,\\
X_{\m{inj}} & = -\frac{4 G \omega_m \sqrt{\kappa}}{ \kappa - 2 i \omega } \chi_{\m{fb}} \,,
\end{split}
\end{equation}
with $\chi_{\m{fb}}$ given by \eref{eq:chifb}.


\subsection{Noise considerations}
In this section, we assume that the optimal feedback condition is met, that is, $\phi = \phi_m$ [\eref{eq:phimax}]. In this case,
\begin{multline}
f_{\rm fb}[\omega] =  \frac{i2GA_0}{\sqrt{\kappa^2/4+\omega_m^2}}x[\omega]\\+ \frac{A_0\omega_m}{\sqrt{\kappa}}e^{-i\phi\frac{\omega}{\omega_m}}\Big((\kappa\chi_c[\omega]-1)y_{c, {\rm in}}[\omega]+y_{\rm add}\Big).
\end{multline}
The first term of the feedback force is responsible for feedback damping. The second term accounts for noise coming from the cavity quantum fluctuations re-injected in the feedback loop, as well as added noise from the detection stage (mainly amplifier added noise), also fed back to the oscillator. As a result, the position satisfies the equation
\begin{multline}
\label{allnoises}
x = \chi_{\rm fb}  \Big[ f_{\rm th} - 2 \omega_m G x_c \\
+ \frac{A_0\omega_m}{\sqrt{\kappa}}e^{-i\phi\frac{\omega}{\omega_m}}\Big((\kappa\chi_c-1)y_{c, {\rm in}}+y_{\rm add}\Big) \Big]
\end{multline}
with $x_c = \chi_c  \sqrt{\kappa} x_\mathrm{c,in}$. In the equation above, the first term accounts for thermal noise, the second accounts for measurement backaction noise, and the third for the total measurement noise fed back to the oscillator, which is a sum of the two contributions discussed above. All noise processes appearing in these equations are uncorrelated, so that contributions do not interfere together. Assuming (as always) that $\gamma\ll\kappa$ and that the cavity is in its ground state, the backaction noise arising from the second term in equation \eref{allnoises} gives \eref{eq:nqba}. The third term in equation \eref{allnoises} leads to a total measurement noise fed back to the oscillator, \eref{eq:nFB}.

\section{Spectrum in the unresolved sideband situation}
\label{sec:badcavityresults}

The ``noise squashing" in the bad-cavity situation allows for simple analytical results. At the optimum feedback phase in this situation, $\phi_m = \pi/2$, the in-loop heterodyne spectrum can be understood as consisting of two Lorentzians with opposite signs:
\begin{equation}
\label{eq:squashbadcav}
\begin{split}
S_{\m{out,x}}[\omega] & = \frac{8 G^2}{\kappa} S_x[\omega]\,, \\
S_{\m{out,\pm}}[\omega] & = - \frac{G A_0 \omega_m \gamma_{\m{eff}}/\kappa (n_{\m{add}} + \puoli)}{(\omega \mp \omega_m)^2 + (\frac{\gamma_{\m{eff}}}{2})^2 } \,, \\
\end{split}
\end{equation}
and $S_{\m{out}}[\omega] = S_{\m{out,x}}[\omega] + S_{\m{out,-}}[\omega]+ S_{\m{out,+}}[\omega] + n_{\m{add}} + \puoli$. The term $S_{\m{out,x}}[\omega]$ gives the sideband asymmetry, while the squashing term is the same for both lower and upper sidebands.

In the case of weak feedback, $S_{\m{out,\pm}}$ are negligible.

\section{Electromechanical forces}
\label{app:emforce}

The intracavity field annihilation operator can be decomposed into $a=\alpha(t)+\alpha_f(t)+\tilde{a}(t)$, where $\tilde{a}$ is a (quantum) annihilation operator associated with a small fluctuation, $\alpha$ is the classical complex amplitude of the probe tone oscillating at $\omega_c$ or $\omega_c+\omega_m$, and $\alpha_f$ is the classical complex amplitude of the feedback tone oscillating at $\omega_c+\Delta_f\pm \omega_m$. 
The original non-linear evolution equation for the mechanical oscillator momentum is:
%
\begin{equation}
\begin{array}{*3{>{\displaystyle}l}}
    \dot{p}&=&-\gamma p - \omega_{m} x - \sqrt{2} g_0 a^\dagger a + \frac{f_{\rm th}}{\omega_m} \,.
\end{array}
\end{equation}
The total electromechanical force $- \sqrt{2} g_0 a^\dagger a$ contains the following terms:
\begin{itemize}
    \item $-\sqrt{2}g_0 |\alpha_f|^2$: feedback force
    \item $-\sqrt{2}g_0 |\alpha|^2$: a DC force from the probe tone
    \item $-\sqrt{2}g_0 (\alpha \tilde{a}^\dagger +\alpha^* \tilde{a})$: dynamical and quantum backaction from the probe tone
    \item $-\sqrt{2}g_0 (\alpha_f \tilde{a}^\dagger +\alpha_f^* \tilde{a}) $: dynamical and quantum backaction from the feedback tone
\end{itemize}
 The force also contains cross-products of the probe and feedback tones (resonating at $\Delta_f$ or $\Delta_f+\omega_m$) that are largely out of resonance with the mechanical oscillator and whose impact is neglected. Finally, it also contains terms such as $-2g_0\tilde{a}^\dagger\tilde{a}$ coming from quantum fluctuations of the cavity field, which is, as usual in driven cavity optomechanics, neglected compared to all other forces as it is typically much weaker.

\subsection{Probe tone at the cavity center}

We now derive $\alpha_f$ to give the expression of the feedback force, in the situation where the probe tone is sent at the cavity center. The position signal in frequency space is 
\begin{equation}
    x[\omega]\equiv \frac{b[\omega] + b^\dagger[\omega]}{\sqrt{2}} \,,
\end{equation}
with $b$ ($b^\dagger$) the annihilation (creation) operator in the laboratory frame.
In the following, we will also denote phase-shifted position signals 
\begin{equation}
x_{\varphi}[\omega]\equiv \frac{b[\omega]e^{i\varphi} + b^\dagger[\omega] e^{-i\varphi}}{\sqrt{2}}
\end{equation}
The inverse Fourier transform of $x_{\varphi}$ is, in the limit of small delays compared to the decoherence time, the delayed position signal $x_\varphi(t)\simeq x(t-\varphi/\omega_m)$. In the following paragraph, we will assume the small delay condition systematically satisfied. The cavity quadrature coupled to the motion is:
\begin{equation}
y_{c}[\omega] = \chi_c[\omega]\bigg[-\sqrt{2}G \,\Big(b[\omega]+b^\dagger[\omega]\Big)\,    + \sqrt{\kappa} \,y_{c,\, \rm in}[\omega]\bigg]
\end{equation}
Each operator ($b$, $b^\dagger$) selects a narrow frequency range ($+\omega_m$, $-\omega_m$) over which the cavity susceptibility can be considered constant. That is, $b[\omega]$ and $b^\dagger[\omega]$ sample the cavity susceptibility at different (opposite) frequencies $\pm\omega_m$, such that:
\begin{multline}
y_{c}[\omega] \simeq  -\sqrt{2}G|\chi_c[\omega_m]| \,\Big(b[\omega]e^{i\phi_0}+b^\dagger[\omega]e^{-i\phi_0}\Big)\\   +\chi_c[\omega] \sqrt{\kappa} \,y_{c,\,\rm in}[\omega]
\end{multline}
where 
\begin{equation}
\begin{array}{*3{>{\displaystyle}l}}
\phi_{\rm 0} &\equiv& {\rm arg}\{\chi_c[\omega_m]\} = \arctan \Big(\frac{2\omega_m}{\kappa}\Big)\\[8pt]
|\chi_c[\omega_m]| &\equiv& (\kappa^2/4+\omega_m^2)^{-1/2}.
\end{array}
\end{equation}
Neglecting cavity noise, the corresponding output quadrature is:
\begin{equation}
\begin{array}{*3{>{\displaystyle}l}}
    y_{\rm out}[\omega]&=& \sqrt{\kappa} y_c[\omega]-y_{c,\, \rm in}[\omega]\\[5pt]
    &=& \frac{2\sqrt{\kappa }G}{\sqrt{\kappa^2/4+\omega_m^2}} \,x_{\phi_0}[\omega] 
     + (\kappa \chi_c[\omega]-1)\,y_{c,\, \rm in}[\omega]\,.
    \end{array}
\end{equation}
We neglect cavity noise in the following. The feedback loop applies a filter which amplifies this signal by a gain $\mathcal{G}$
and delays it by $\tau$. The equivalent phase shift for an oscillator at $\omega_m$ is $\phi_\tau\equiv\omega_m\tau$. The unit of $\mathcal{G}$ is chosen such that the result of the filtering operation is a dimensionless signal $s(t)$ whose Fourier transform is
\begin{equation}
s[\omega]=  \frac{2\mathcal{G}\sqrt{\kappa }G}{\sqrt{\kappa^2/4+\omega_m^2}}  \,x_{\phi_0+\phi_\tau}[\omega] \,.
\end{equation}
This signal is sent to a phase modulator driven by a coherent pump at $\omega_f$ and of amplitude $\alpha_0/\pi$. The result of this modulation is an electronic signal
\begin{equation}
    \alpha_{f, \rm mod}(t) = \alpha_0/\pi \sin(\omega_f t+s(t)).
\end{equation}
In the equation above, the conversion factor affecting $s(t)$ in the mixing operation is also integrated into $\mathcal{G}$, and therefore $s(t)$, for convenience. Finally, if $|s(t)|\ll 2\pi$, using $\sin s(t)\simeq s(t)$ and $\cos s(t)\simeq 1$, this signal corresponds to the sum of a strong coherent tone oscillating at $\omega_f$ and of a weaker signal at $\omega_f$ whose amplitude is modulated by $s(t)$:
\begin{equation}
    \alpha_{f, \rm mod}(t) \simeq \alpha_0/\pi\Big(\sin \omega_f t + s(t)\cos\omega_f t\Big)\,.
\end{equation}
The corresponding complex amplitude in the frame oscillating at the cavity resonance frequency is
\begin{equation}
    \alpha_{f, \rm in}[\omega]= i\alpha_0\delta(\omega-\Delta_f) + \frac{\alpha_0\mathcal{G}\sqrt{\kappa }2G}{\sqrt{\kappa^2/4+\omega_m^2}}  \,x_{\phi_0+\phi_\tau}[\omega-\Delta_f] \,.
\end{equation}
This signal is driving the cavity to apply the feedback force. The spectrum of this signal is a triplet of peaks: one strong peak of amplitude $\alpha_0$ at $\omega=\omega_c+\Delta_f$ and two weaker sidebands (weak as per the assumption above $|s(t)|\ll 2\pi$), whose spectrum reproduces the spectrum of $x$, oscillating at frequencies $\omega = \omega_c+\Delta_f\pm \omega_m$. Since these three peaks sit at significantly different frequencies compared to the cavity linewidth, they are affected differently by the cavity susceptibility. In particular, they will be affected by different phase shifts due to the cavity susceptibility denoted $\phi_1$, $\phi_+$ and $\phi_-$.
\begin{multline}
\alpha_{f}[\omega]=  \frac{\alpha_0\mathcal{G}\kappa \sqrt{2}G}{\sqrt{\kappa^2/4+\omega_m^2}}\Bigg(\frac{b[\omega-\Delta_f]e^{i(\phi_0+\phi_\tau+\phi_{+})}}{\sqrt{\kappa^2/4 + (\Delta_f+\omega_m)^2}}\\[15pt]
+\frac{b^\dagger[\omega-\Delta_f]e^{-i(\phi_0+\phi_\tau-\phi_{-})}}{\sqrt{\kappa^2/4 + (\Delta_f-\omega_m)^2}}\Bigg)
+\alpha_0\sqrt{\kappa}\, \frac{\delta(\omega-\Delta_f)e^{i(\phi_{1}+\pi/2)}}{\sqrt{\kappa^2/4+\Delta_f^2}}
\end{multline}
 with
 \begin{multline}
     \phi_1 = \tan^{-1}\frac{2\Delta_f}{\kappa},\\
     \phi_+ = \tan^{-1}\frac{2(\Delta_f+\omega_m)}{\kappa}\,,\;\text{and }  \phi_- = \tan^{-1}\frac{2(\Delta_f-\omega_m)}{\kappa}\,.
 \end{multline}

\subsubsection{Feedback force}

The linear feedback force therefore comes from a cross-product of either sideband of the drive with the central peak of the drive (neglecting the added noise contribution for now):
\begin{multline}
    f_{\rm fb}[\omega]  =- g_0\frac{\kappa^{3/2}\alpha_0^2 \mathcal{G}2G}{\sqrt{(\kappa^2/4+\omega_m^2)(\kappa^2/4+\Delta_f^2)}} \\ \Bigg(\frac{x_{\phi_0+\phi_\tau+\phi_{+}-\phi_1-\pi/2}[\omega]}{\sqrt{\kappa^2/4 + (\Delta_f+\omega_m)^2}}\\
    +  \frac{x_{\phi_0+\phi_\tau-\phi_{-}+\phi_1+\pi/2}[\omega]}{\sqrt{\kappa^2/4 + (\Delta_f-\omega_m)^2}}\Bigg)
\end{multline}
The force therefore has two contributions. This results from each of the sideband of the driving triplet interfering with the central peak to give a force contribution. Each of these force contains a phase-shifted version of the position, with a different phase-shift for both, due to the frequency-dependence of the cavity susceptibility phase. In the limit of small delays compared to the decoherence rate, the sum of two position signals with different phase shifts is itself a phase-shifted position signal. Denoting the phase shifts:  $\varphi= \phi_\tau+\phi_{+}-\phi_1-\pi/2$ and $\varphi'= \phi_\tau-\phi_{-}+\phi_1+\pi/2$, and the coefficients $A=\frac{\kappa}{\sqrt{\kappa^2/4+(\Delta_f+\omega_m)^2}}$, $B=\frac{\kappa}{\sqrt{\kappa^2/4+(\Delta_f-\omega_m)^2}}$, the force is then written (again neglecting the noise contribution)
\begin{equation}
    f_{\rm fb}[\omega] =  -\frac{2G}{\sqrt{\kappa^2/4+\omega_m^2}} \frac{g_0\sqrt{\kappa}\alpha_0^2 \mathcal{G}}{\sqrt{\kappa^2/4+\Delta_f^2}} \,D\, x_{\phi_0+\phi}[\omega]
\end{equation}
with $D\equiv\sqrt{A^2+B^2+2AB\cos(\varphi-\varphi')}$ and  $\phi \equiv {\rm arctan} \frac{A\sin\varphi+B\sin\varphi'}{A\cos\varphi+B\cos\varphi'}$. We can now identify the phase shift $\phi$ of the feedback filtering application as defined in the main text, as well as the gain $A_0$
\begin{equation}
    A_0 = \frac{g_0\sqrt{\kappa}\alpha_0^2 \mathcal{G}}{\sqrt{\kappa^2/4+\Delta_f^2}} D
\end{equation}
proportional to parameter $\mathcal{G}$.
We also see that, because the two position signals contributing to the feedback force are not affected by the same phase-shift, their sum is only partially constructive. Indeed, the gain $A_0$ is maximum when $\varphi$ and $\varphi'$ are equal. On the other hand, if one force contribution largely dominates  over the other, then the interference between these two contributions is weak. The only very unfavorable situation, which makes $A_0$ vanish, is therefore the very bad cavity limit, wherein the amplitudes of the two contributions are similar and $\varphi-\varphi'\simeq -\pi$. In the experimental situation, not accounting for phase shifts incurred in the propagation in transmission lines, the phase difference between force components  $\varphi-\varphi' = \phi_{+}+\phi_{-}-2\phi_1-\pi$ is about $-176^\circ$, which would be quite unfavorable. However, the sideband at $\omega_c+\omega_m+\Delta_f$ of the feedback triplet is more than twice stronger than the sideband at $\omega_c-\omega_m+\Delta_f$, limiting the destructive interference's impact. Furthermore, as they come from signals at different frequencies, $\varphi$ and $\varphi'$ are also impacted by different phase shifts accumulated in the propagation in transmission lines, which are not accounted for in the latter estimate. In conclusion, even in the worst phase configuration where the two feedback force contributions interfere destructively, the fact that one dominates over the other ensures that the destructive interference is not complete and that the feedback force is significant.

\section{Data analysis}
\label{app:calibrations}

The effective coupling $G$ at a given generator power is obtained based on a standard power sweep in a sideband cooling measurement, where the pump frequency is set at the red sideband. The optomechanical damping is fitted linearly with the generator power $P$:
\begin{equation}
\gamma_{\m{opt}} = \mathcal{P} P \,,
\end{equation}
where $\mathcal{P}$ is the calibration coefficient. 
The effective coupling under the red-sideband probing is obtained from \eref{eq:gammaopt} as
\begin{equation}
G_{\m{rsb}} = \puoli \sqrt{\mathcal{P} P \kappa \LL[ 1+ \LL(\frac{\kappa}{4 \omega_m} \RR)^2 \RR] }\,.
\end{equation}

In determining $G$ in the feedback experiment, we have to account for the specific probe tone detuning because the field amplitude in the cavity, at a given generator power, depends on the cavity susceptibility:
\begin{equation}
G = \frac{|\chi_c(\Delta)|}{|\chi_c(-\omega_m)|} G_{\m{rsb}} \,,
\end{equation}
where $\Delta$ is the specific detuning, which can be either $\Delta = 0$, or $\Delta = \omega_m$.


The gain $A_0$ utilized in the theoretical discussion is not a quantity directly applicable to the experiment. The experimentally relevant gain value $\mathcal{G}$ is simply proportional to it, with an unknown coefficient coming from the transduction.

The values of $A_0$ calibrated as described immediately below are used when fitting the theoretically obtained output spectrum to the feedback data, and to infer the mechanical occupation.

\subsection{Resonant probing}

We fit the measured linewidths as follows to obtain the calibration coefficient $\mathcal{L}$:
\begin{equation}
\gamma_{\m{fb}} = \mathcal{L} g \,.
\label{eq:Lcal}
\end{equation}

We combine \eref{eq:Lcal} with \eref{eq:gammafb}:
\begin{equation}
A_0 = \frac{\mathcal{L} \mathcal{G} \sqrt{\kappa^2 + 4\omega_m^2}}{4 G} \,.
\label{eq:Lcaluse}
\end{equation}

\subsection{Blue-sideband probing}

Here, we use the damping stated in \eref{eq:gammaeffBSB}, and obtain at the optimum phase
\begin{equation}
\label{eq:gammaeffBSB2}
\gamma_{\m{eff}} = \gamma - \gamma_{\m{opt}} + \gamma_{\m{fb,bsb}} \,,
\end{equation}
where
\begin{equation}
\label{eq:gammafbBSB}
 \gamma_{\m{fb,bsb}} = \frac{4 A_0 G \sqrt{\kappa^4+20 \kappa^2 \omega_m^2+64 \omega_m^4}}{\kappa^3+16 \kappa \omega_m^2} \,.
\end{equation}

We fit to Eqs.~(\ref{eq:gammaeffBSB2},\ref{eq:gammafbBSB}) the measured linewidth with increasing gain, in a similar manner as described in \eref{eq:Lcaluse} for the resonant probing situation.

\end{document}